\documentclass[twocolumn]{aastex7}

\usepackage{amsmath}
\usepackage{bbm,bm}
\usepackage{multirow}
\usepackage{colortbl}
\usepackage{hyperref}
\usepackage{outlines}
\usepackage{comment,color}

\graphicspath{{./}{figures/}}

\shorttitle{}
\shortauthors{}

\begin{document}

\title{Estimation of Magnetic Fields from Synchrotron Emission: Numerical Tests}

\author[0000-0001-8840-2538]{Nora B. Linzer}
\affiliation{Department of Astrophysical Sciences, Princeton University, 4 Ivy Lane, Princeton, NJ 08544, USA}
\email{nlinzer@princeton.edu}

\author[0000-0002-5708-1927]{Lucia Armillotta}
\affiliation{INAF Arcetri Astrophysical Observatory, Largo Enrico Fermi 5, Firenze, 50125, Italy}
\affiliation{Department of Astrophysical Sciences, Princeton University, 4 Ivy Lane, Princeton, NJ 08544, USA}
\email{lucia.armillotta@princeton.edu}

\author[0000-0002-0509-9113]{Eve C. Ostriker}
\affiliation{Department of Astrophysical Sciences, Princeton University, 4 Ivy Lane, Princeton, NJ 08544, USA}
\affiliation{Institute for Advanced Study, 1 Einstein Drive, Princeton, NJ 08540, USA}
\email{eco@astro.princeton.edu}

\author[0000-0001-9185-5044]{Eliot Quataert}
\affiliation{Department of Astrophysical Sciences, Princeton University, 4 Ivy Lane, Princeton, NJ 08544, USA}
\email{quataert@princeton.edu}

\begin{abstract}
We use models of spectrally resolved cosmic ray (CR) transport in TIGRESS MHD simulations of the local ISM to produce synthetic synchrotron emission and to test, on scales from a few kpc down to $\sim10$ pc, the traditional estimate of magnetic field strength based on the assumption of equipartition between the magnetic and total CR energy densities. Our analysis shows that the traditional equipartition estimate works well at the kpc scale of the simulation box, but breaks down at smaller scales. We find that the predicted magnetic field strength can be improved at small scales by assuming a constant CR energy density across each mock radio observation. The large-scale mean CR energy density can be estimated by assuming equipartition with the large-scale mean magnetic energy density, or as a function of additional observable quantities such as the star formation rate surface density or gas weight. In addition to estimating the magnetic field strength, we use synthetic polarized emission to create maps of the magnetic field direction. We find that the true magnetic field direction can be recovered well from the mock observations.
\end{abstract}
\keywords{}

\section{Introduction}\label{sec:intro}
\setcounter{footnote}{0}

Synchrotron radiation, produced through the interaction of cosmic ray electrons (CREs) with magnetic fields, is the most important probe of the magnetic field strength and structure in extragalactic sources \citep[e.g.][]{beck_magnetic_2016, han_observing_2017}. Other measures of magnetic field strength, such as Zeeman splitting and Faraday rotation, are feasible in the Milky Way but are more challenging in distant galaxies \citep[e.g.][]{heiles_magnetic_2005,Robishaw2008, han_observing_2017}. To extract information about the magnetic field from radio synchrotron emission, however, requires assumptions about the CRE and cosmic ray (CR) proton spectra and the relative CR and magnetic field spatial distributions. Since the overall level of synchrotron emission depends on both magnetic and CR energy, either additional information regarding one or the other energy density or an additional assumption is needed for a joint constraint.  

The most common simplifying assumption is that of equipartition between the magnetic and total CR energy densities \citep[e.g.][]{govoni_magnetic_2004}. In the solar neighborhood, it has been shown through independent measurements that these two values are roughly equal, each approximately 1 eV cm$^{-3}$ \citep[e.g.][]{heiles_magnetic_2005, zweibel_basis_2017}. Additionally, equipartition has been considered to be a reasonable assumption because the CRs are confined by the magnetized interstellar medium (ISM), and the magnetic and CR energy densities could therefore be coupled \citep[e.g.][]{draine_physics_2011}.\footnote{One could argue, for example, that if horizontal magnetic field lines need to be forced open by CR pressure in order for the CRs to escape vertically from a galactic disk, then the CR pressure must build up until (rough) equipartition is achieved.} There is not a clear justification, however, for why these energy densities should necessarily be equal in all environments and across all physical scales. Below $\sim 1$ kpc, for example, the CR and magnetic field energy densities are not likely to be in equipartition due to the long mean free path of CR propagation in comparison to much smaller scale ISM dynamical processes that create structure in the magnetic field \citep[e.g.][]{stepanov_observational_2014, seta2018, seta_revisiting_2019}.

Standard equipartition magnetic field estimates also rely on assumptions about the local CR spectrum \citep[e.g.][]{beck_revised_2005}. Although synchrotron emission is primarily produced through interactions of CREs with the magnetic field, the total CR energy density is dominated  by protons with energies of approximately 1 GeV \citep[e.g.][]{zweibel_microphysics_2013}. In the solar neighborhood, there are direct measurements of the relative normalization and spectral shape of the proton and electron spectra, but for extragalactic sources these values typically must be assumed (except to some extent in the few sources with both gamma-ray and radio observations; e.g., \citealt{McDaniel2019}). The choice of CRE spectral index has been shown to make an impact on the resulting measurement of magnetic field strength \citep{padovani_spectral_2021, Bracco+24}.

Other groups have previously used simulations to test the equipartition assumption. On the one hand, \cite{2024MNRAS.52711707P} use magnetohydrodynamic (MHD) cosmological zoom-in FIRE simulations \citep{hopkins_first_2022}, which self-consistently evolve CRs over a range of energies, to produce mock synchrotron observations and estimate the equipartition magnetic field strength. They find this method to underestimate the true magnetic field and present a possible improvement to the method by considering the volume filling fraction of synchrotron emission in the multi-phase ISM. On the other hand, \cite{dacunha_overestimation_2024} impose different CR distributions, for instance fixed equipartition with the simulated magnetic energy density or an exponentially decreasing distribution of CR energy as a function of galactic height and radius, onto galaxy simulations that do not self-consistently evolve CRs to produce mock synchrotron emission. Using these synthetic observations, they find that the equipartition assumption generally overestimates the magnetic field strength with some dependence on the CR power-law index and the underlying field.

In this paper, we test different methods, including the conventional equipartition approach, for estimating the varying magnetic field strength on large ($\gtrsim$ 1 kpc) and small ($\sim 10$ pc) scales in the ISM from synthetic synchrotron observations. To do so, we make use of the TIGRESS MHD simulations of the magnetized, multiphase ISM with star formation and feedback, post-processed with a two-moment solver for the transport of spectrally resolved CR protons and electrons, as described in \cite{Linzer2025,Armillotta2025}. These simulations reproduce the CR spectra directly observed in the solar neighborhood and have accurate magnetic field structure. As we have information on the simulated magnetic field direction, we can also produce synthetic polarized emission and test the extent to which observations of synchrotron emission can be used to correctly infer the magnetic field structure. 

In \autoref{sec:methods} we summarize the TIGRESS models and the CR transport method. We also describe the method to produce synthetic synchrotron emission from which we estimate the equipartition magnetic field strength. In \autoref{sec:results}, we present our results including our test of the equipartition estimate and possible alternative methods to extract the magnetic field strength with additional observations. We conclude with a discussion of polarized synchrotron emission, considering how well the magnetic field direction can be recovered from observations of the polarization angle (\autoref{sec:pol}). In \autoref{sec:disc} we discuss implications of our results on the use of the equipartition estimate, particularly its application at small scales. Finally, we summarize our results and present our conclusions in \autoref{sec:conc}.

\section{Methods}\label{sec:methods}

First, we describe the CR transport method implemented by \cite{armillotta_cosmic-ray_2021} and \cite{Linzer2025} to post process the TIGRESS simulations of the multiphase ISM \citep{kim_three-phase_2017}. We then summarize the method used to produce synthetic maps of synchrotron emission as detailed in \cite{padovani_spectral_2021}. Finally, we review the equipartition estimate of magnetic fields as derived in \cite{beck_revised_2005}.

\subsection{CR transport in TIGRESS}

For our analysis, we use the MHD TIGRESS simulations of the multiphase ISM post-processed with a solver for transport of spectrally resolved CRs. TIGRESS \citep{kim_three-phase_2017} simulates kpc-scale patches of the ISM using the \textit{Athena} code \citep{stone_athena_2008}. The TIGRESS framework includes physics as needed to establish a self-consistent quasi-equilibrium ISM state, including star formation and supernova feedback, photoelectric heating of warm and cool gas (scaled by the star formation rate), cooling mechanisms for the full range of ISM phases, and both self- and external gravity. These simulations produce an accurate representation of the magnetic field distribution, which is created by the interaction of turbulence and large-scale fountain flows/outflows (both primarily driven by supernovae), sheared galactic rotation, and other effects such as differential buoyancy of multiphase gas in the gravitational field.  This magnetic field is important for modeling both CR transport and synchrotron emission. For the full details of the TIGRESS numerical framework, see \cite{kim_three-phase_2017} and \citet{kim_first_2020}. We note that CRs were not modeled in the original TIGRESS simulations that we use here, so the full dynamical back-reaction of CRs on the gas was not included. In recent tests, however (C.-G. Kim et al, in prep), we find that only extraplanar gas flows are altered by CR forces, while the midplane regions considered here are not affected.  

The TIGRESS models we use in this work represent solar neighborhood conditions, although there are many other models which span a wide range of galactic conditions, most recently allowing for full adaptive ray tracing radiation and a range of metallicity \citep{2024ApJ...972...67K,Kim_2024}. Here, we consider post-processed snapshots from both the {\tt R8} and the {\tt R8 Arm} model, from \citet{kim_first_2020} and \citet{kim_arm_2020}, respectively. The {\tt R8} model has a box size of $L_x = L_y = 1024$ pc and $L_z = 7168$ pc with a resolution of $\Delta x = 8$ pc. We include eight snapshots from this model in our analysis, the same presented in detail in \cite{Linzer2025}. One goal of this work is to test the dependence of the equipartition estimate on physical scale. Therefore, we additionally post-process four snapshots from the {\tt R8 Arm} model. This model has a larger box size compared to {\tt R8}, and includes a denser, central arm feature. The {\tt R8 Arm} box has $L_x = 3141.6$ pc and $L_y = L_z = 6283.2$ pc with a resolution of $\Delta x \approx 12.3$ pc. \citet{2025arXiv250903519H} analyzes CR transport in the {\tt R8 Arm} model in more detail, along with TIGRESS models representing different galactic environments beyond the solar neighborhood.

CR transport is solved numerically using the two-moment method described in \cite{jiang_new_2018} and later adapted by \cite{armillotta_cosmic-ray_2021} to model a single CR fluid made of protons with kinetic energy of approximately 1 GeV. In \cite{Linzer2025}, we described our extension of this approach to track the transport of multiple CR fluid components, each representing either protons or electrons within a specific range of momenta. Each TIGRESS snapshot is post-processed with this scheme modeling ten CR momentum components, five each of protons and electrons, with kinetic energies between approximately 1-100 GeV. 

The CR transport framework includes streaming, advection, and diffusion with  spatial- and energy-dependent scattering coefficients determined self-consistently by balancing streaming-driven Alfvén wave excitation with damping mediated by local gas properties. The CRs undergo energy-dependent losses of multiple types including synchrotron, inverse Compton, and ionization. There are two major caveats to this method, as described in more detail in \cite{Linzer2025} and \cite{Armillotta2025}. The first is that the CR momentum bins are evolved independently without allowing for energy transfer between them. Despite this limitation, however, we do recover the directly observed CR spectra, both of protons and electrons, in the solar neighborhood. Secondly, we do not self-consistently evolve the CRs with the gas. However, because the CR pressure is quite uniform in neutral gas where waves are strongly damped, the CRs would not in any case have a significant dynamical impact on the ISM structure in the midplane, where the majority of synchrotron emission is produced (see also \citealt{Rathjen2023, sike2024cosmic, Thomas2025} and C.-G. Kim et al., in prep.).

In our post-processing scheme, we run the simulations until the CR energy density in every energy bin reaches a steady state. During this stage, the gas is frozen and there is no MHD evolution. In the {\tt R8} models, we follow this step with a short period of MHD relaxation in which the gas and magnetic field are free to evolve in reaction to the CRs. All results presented for the {\tt R8} model are after the MHD relaxation, although we find that the synthetic synchrotron emission, and the resulting equipartition estimate of the magnetic field, does not change significantly before and after the MHD is allowed to evolve. The {\tt R8 Arm} post-processing does not include this MHD relaxation step due to its more complex boundary conditions, which would necessitate modifications to the code.

\subsection{Synchrotron emission}
\label{sec:sync}

To generate maps of synthetic synchrotron emission from our TIGRESS snapshots, we use the method described by \cite{padovani_spectral_2021} which they summarize from the results of \cite{1965ARA&A...3..297G}. We present the relevant equations here, and point to \cite{padovani_spectral_2021} for the full description.

The synchrotron specific emissivity has two components, polarized either parallel or perpendicular to the magnetic field projected along the line of sight. These are given by
\begin{equation}
    \epsilon_{\nu,\parallel} = \int_{m_e c^2}^\infty \frac{j_e(E)}{v_e(E)}P_{\nu,\parallel}(E, B_\perp)dE,
    \label{eq:eps_par}
\end{equation}
\begin{equation}
    \epsilon_{\nu,\perp} = \int_{m_e c^2}^\infty \frac{j_e(E)}{v_e(E)}P_{\nu,\perp}(E, B_\perp)dE.
    \label{eq:eps_perp}
\end{equation}
Here, the CRE spectrum $j_e(E)$ is computed as follows. First, we calculate the value of $j_\mathrm{e}$ at the central energy $E_j$ of each bin $j$ from its energy density $e_{\mathrm{c},j}$ (a direct simulation outcome) using $j_{\mathrm{e},j}(E_j) \approx e_{\mathrm{c},j} v_{\mathrm{e},j}/(4 \pi  E_j^2 d \mathrm{ln}E)$ where $v_{\mathrm{e},j}$ is the velocity of the CREs in bin $j$. This approximation holds due to the sufficiently small bin width $d \mathrm{ln}E = 0.1$. Next, we interpolate the spectrum between two neighboring bins with a power-law function. Additionally, we extrapolate the spectrum to CR energies of $1-10^3$ GeV assuming a constant power law slope beyond the simulated energy bins. This assumption is consistent with direct observations in the solar neighborhood \citep[e.g.][]{padovani_spectral_2021}.

$P_\nu$ is the synchrotron power and is a function of frequency ($\nu$), CR energy ($E$), and magnetic field strength perpendicular to the line of sight ($B_\perp$). For the full expression see e.g. Equation 19.29 in \cite{shu_radiation} or Equation 2 in \cite{padovani_spectral_2021}. The CR velocity, $v_e$, is approximately $c$ for our relativistic electrons.

These emissivities are integrated along the line of sight (the $z$ direction, for our assumption of a face-on disk) to find the specific intensity, $I_\nu$, along with the Stokes parameters, $Q_\nu$ and $U_\nu$, as follows:
\begin{equation}
    I_\nu = \int_{-L_z/2}^{+L_z/2}(\epsilon_{\nu,\parallel} + \epsilon_{\nu,\perp})dz,
    \label{eq:Iv_synch}
\end{equation}
\begin{equation}
    Q_\nu = \int_{-L_z/2}^{+L_z/2}(\epsilon_{\nu,\perp} - \epsilon_{\nu,\parallel})\textrm{cos}(2\varphi) dz,
\end{equation}
\begin{equation}
    U_\nu = \int_{-L_z/2}^{+L_z/2}(\epsilon_{\nu,\perp} - \epsilon_{\nu,\parallel})\textrm{sin}(2\varphi) dz.
\end{equation}
In these equations, $\varphi$ describes the local polarization angle, which is found by rotating $90^\circ$ relative to $B_\perp$ in the observed plane.

The pixel scale for the mock observables is 8 pc for the {\tt R8} model and 12.3 pc for the {\tt R8 Arm} model, the same as the simulation resolution in each case. The simulation resolution is higher than many current observations \citep[e.g.][]{mora-partiarroyo_chang-es_2019, stein_chang-es_2023}, but is comparable to the Local Group L-band Survey \citep[LGLBS,][]{koch2025} or to future surveys such as the next-generation VLA \citep[e.g.][]{mckinnon_2019} and Square Kilometre Array \citep[e.g.][]{braun_2015}.

From these mock observables, we can obtain the polarization angle as
\begin{equation}
    \varphi = \frac{1}{2}\textrm{arctan}\left(\frac{U_\nu}{Q_\nu}\right).
    \label{eq:phi}
\end{equation}

\subsection{Equipartition estimate}
\label{sec:eq_def}

Using our mock synchrotron observations, we will test the equipartition estimate of magnetic field strength as presented in \cite{beck_revised_2005}, which we will refer to as the traditional equipartition estimate. We summarize the results relevant to our analysis here, and point to Appendix A in \cite{beck_revised_2005} for the full derivation.

The distribution of CR protons is represented by a broken power law with number density given by 

\begin{equation}
n_{\rm p}(E) \textrm{d}E =
    \begin{cases}
        n_{\rm p, 0} (E_p/E_0)^{-\gamma} \textrm{d}E & \text{if } E < E_p  \\
        n_{\rm p, 0} (E/E_0)^{-\gamma} \textrm{d}E & \text{if } E > E_p 
    \end{cases}
    \label{eq:n_p}
\end{equation}
where $n_{\rm p, 0}$ represents the normalization at reference energy $E_0$ and $E_p$ is the rest mass energy of protons. 

At energies greater than $E_p$, and below the energy at which synchrotron and IC losses become significant ($E_{\rm syn}$), the CRE distribution is assumed to be
\begin{equation}
    n_{\rm e}(E) \textrm{d}E = n_{\rm e, 0} (E/E_0)^{-\gamma_{\rm e}} \textrm{d}E
    \label{eq:n_e}
\end{equation}
with spectral index $\gamma_{\rm e}$. This is related to the CRE spectrum in  \autoref{eq:eps_par} and \autoref{eq:eps_perp} as 
\begin{equation}
    j_e(E) \textrm{d}E = \frac{v_e}{4\pi}n_e(E) \textrm{d}E.
\end{equation}

The total synchrotron intensity at frequency $\nu$ can then be estimated as
\begin{equation}
    I_\nu = c_2(\gamma_{\rm e})n_{\rm e, 0} E_0^{\gamma_{\rm e}}(\nu / 2 c_1)^{(1-\gamma_e)/2}  \ell_z B_{\perp}^{(\gamma_{\rm e} + 1)/2}
    \label{eq:Iv}
\end{equation}
assuming a constant magnetic field and CRE spectrum along the line of sight. This differs from \autoref{eq:Iv_synch} in which the synchrotron emissivity varies as a function of the local CR properties and magnetic field strength. The length scale of synchrotron emission, $\ell_z$, is typically taken to be of order $1-2$ kpc when considering emission from a face-on source \citep[e.g.][]{beck_magnetic_2016, 2024MNRAS.52711707P}. We set $\ell_z = 1$ kpc, which is approximately twice the scale height of synchrotron emission in our simulation, but note that the choice of $\ell_z$ does not significantly impact the final estimate. 

The constants $c_1$ and $c_2$ are defined as
\begin{equation}
    c_1 = \frac{3e}{4\pi m_e^3 c^5} \approx 6.3 \times 10^{18} \textrm{ erg}^{-2}\textrm{ s}^{-1}\textrm{ G}^{-1}
\end{equation}
and
\begin{equation}
    c_2 = \frac{\sqrt{3}e^3}{16\pi m_e c^2}\frac{\gamma_{\rm e} + \frac{7}{3}}{\gamma_{\rm e} + 1}\Gamma \left( \frac{3\gamma_{\rm e} - 1}{12} \right) \Gamma \left( \frac{3\gamma_{\rm e} + 7}{12} \right).
\end{equation}
For the observed spectrum at $\sim$ GeV energy with $\gamma_\mathrm{e}\sim 2.7$, $c_2=8.7\times10^{-24} \textrm{ cm}^3 \textrm{ G}$.

\cite{beck_revised_2005} make the assumption that for energies between $E_p$ and $E_{\rm syn}$, the CR proton and electron distributions have the same spectral index, $\gamma = \gamma_{\rm e}$, and the ratio in the normalization of the two distributions is $n_{\rm p, 0}/n_{\rm e, 0} = K_0 = 100$. 

Using these assumptions, they solve  \autoref{eq:Iv} for $n_{\rm e, 0}$ and rewrite the result in terms of the number density of CR protons, $n_{\rm p, 0}$. Then, integrating the distribution function (\autoref{eq:n_p}) to find the total CR energy density ($e_{\rm c}$) gives
\begin{equation}
   e_{\rm c} = \frac{\gamma (K_0+1)  E_p^{2-\gamma} (\nu / 2 c_1)^{(\gamma-1)/2}}{2(\gamma-2)c_2(\gamma)  } \frac{I_\nu}{\ell_z B_{\perp}^{(\gamma+1)/2}}.
   \label{eq:eq_ec}
\end{equation}
Here, the factor of $K_0+1$ represents the contributions from the combined CR proton and electron populations.

To convert $B_{\perp}$ to the total magnitude of the magnetic field, \cite{beck_revised_2005} set 
\begin{equation}
    \label{eq:B_perp}
    B_\perp^{(\gamma+1)/2} = \left[B \textrm{cos}(i)\right]^{(\gamma+1)/2} = c_4(i)B^{(\gamma+1)/2}
\end{equation}
where $i$ is the inclination angle. In this case, $i=0$ ($c_4 = 1$) corresponds to a face-on view which we use for this analysis; this assumes the magnetic field is primarily parallel to the midplane of the galaxy, which is the case. 

At this point, an assumption must be made for the value of $e_{\rm c}$ to solve for $B$. In the traditional approach, this is where the equipartition assumption is invoked, letting $e_{\rm c} = e_\mathrm{mag}\equiv B^2/8\pi$. Then, solving for the magnetic field strength gives the final estimate,
\begin{equation}
    \label{eq:equipartition}
    B_{\rm eq} = \left[ \frac{4 \pi (2\alpha + 1) (K_0 + 1) E_{\rm p}^{1-2\alpha}(\nu/2c_1)^{\alpha}}{(2\alpha - 1)c_2(\alpha)  c_4(i)} \frac{I_\nu}{\ell_z}\right]^\frac{1}{\alpha + 3}.
\end{equation}
Here, $\alpha = (\gamma-1)/2$ and can be observationally estimated from synchrotron intensity measured at two different frequencies ($\nu_1$ and $\nu_2$) as
\begin{equation}
    \alpha = -\frac{{\rm log}(I_{\nu_1}/I_{\nu_2})}{{\rm log}(\nu_1/\nu_2)}\,.
\end{equation}

\section{Results}
\label{sec:results}

As a preliminary, we first summarize our measurements for the CR energy density $e_\mathrm{c}$ compared to the magnetic energy density $e_\mathrm{mag}$ in our simulations. We then turn to the main investigation. First, we present a comparison between the true, simulated magnetic field strength and the equipartition estimate in both our {\tt R8} and {\tt R8 Arm} models. Then, we consider modified estimates of magnetic field strength using alternative measures of the CR energy density. Finally, we determine how well the magnetic field direction can be recovered from polarization measurements.

\subsection{Measured CR and magnetic energy densities}\label{sec:measured_e}

\begin{figure}
    \centering
    \includegraphics[width=\linewidth]{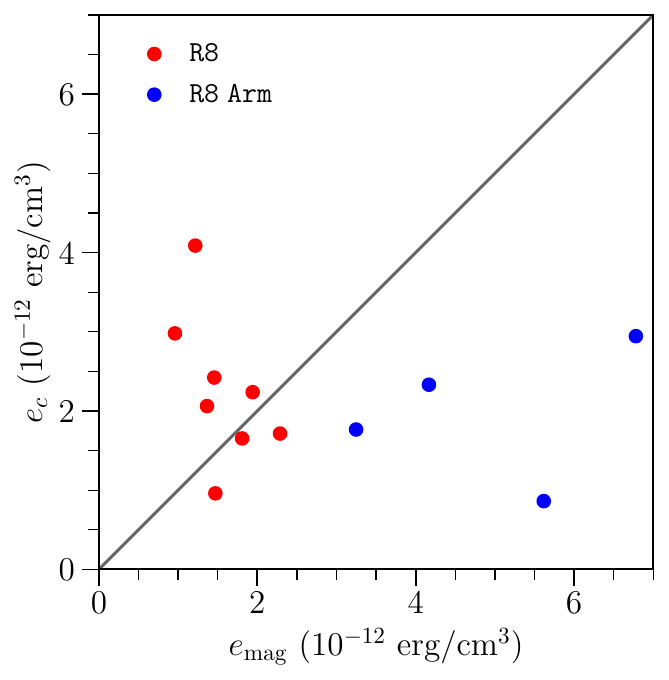}
    \caption{Comparison of the CR and magnetic energy densities. Each point represents the mass-weighted average value for $z < 300$ pc in one time snapshot. This includes both the {\tt R8} and {\tt R8 Arm} models shown in red and blue respectively. The black diagonal line shows $e_c = e_{\rm mag}$.}
    \label{fig:ec_eB}
\end{figure}

First, we directly compare the CR and magnetic energy densities to see their relative magnitude at large scales in our simulation. To do so, we take the mass-weighted average of both of these quantities within the midplane ($z < 300$ pc) of each of our simulation snapshots. The mass-weighted average concentrates on the high-density regions, where the magnetic field is stronger and the associated synchrotron emission is more intense. We show the results for both the {\tt R8} and {\tt R8 Arm} snapshots in \autoref{fig:ec_eB}. 

In the simulation, the total CR energy density above 1 GeV is found by integrating the CR proton spectrum, $j_p$, with respect to CR energy. We define $j_p$ in an analogous way to $j_e$, with the CR proton spectrum in one energy bin given by $j_{\mathrm{p},j}(E_j) \approx e_{\mathrm{c},j} v_{\mathrm{p},j}/(4 \pi  E_j^2 d \mathrm{ln}E)$. We then interpolate between the simulated bins representing $j_p$ as a broken power law. As with $j_e$, we extrapolate $j_p$ to energies of $1-10^3$ GeV assuming a constant power law slope outside of the simulated energies. 

We can see that the CR and magnetic energy densities are of the same order of magnitude in all of the simulation snapshots. However, they are clearly not equal. The value of $e_c/e_{\rm mag}$ varies both between snapshot times within one TIGRESS model and between the different TIGRESS models. Across the {\tt R8} snapshots, we find an average value of $e_c/e_{\rm mag} \approx 1.6$, whereas we find an average of $e_c/e_{\rm mag} \approx 0.4$ for the {\tt R8 Arm} snapshots. Therefore, while equipartition may be a reasonable approximation (within a factor $\sim 2$), it does not hold exactly. 

\subsection{Traditional equipartition estimate} 
\label{sec:eq_est}

\subsubsection{{\tt R8} Model}
\label{sec:R8_eq}

For each of our {\tt R8} snapshots, we produce synthetic maps of synchrotron emission at 1.5 and 3 GHz (\autoref{sec:sync}) corresponding to L and S band observations. Although the absolute magnitude of the predicted magnetic field strength changes slightly depending on the particular modeled frequencies, our final conclusions are insensitive to this choice. From these mock observations, we determine the equipartition magnetic field strength in each pixel of the projected map with \autoref{eq:equipartition}. 

\begin{figure*}
    \centering
    \includegraphics[width=\linewidth]{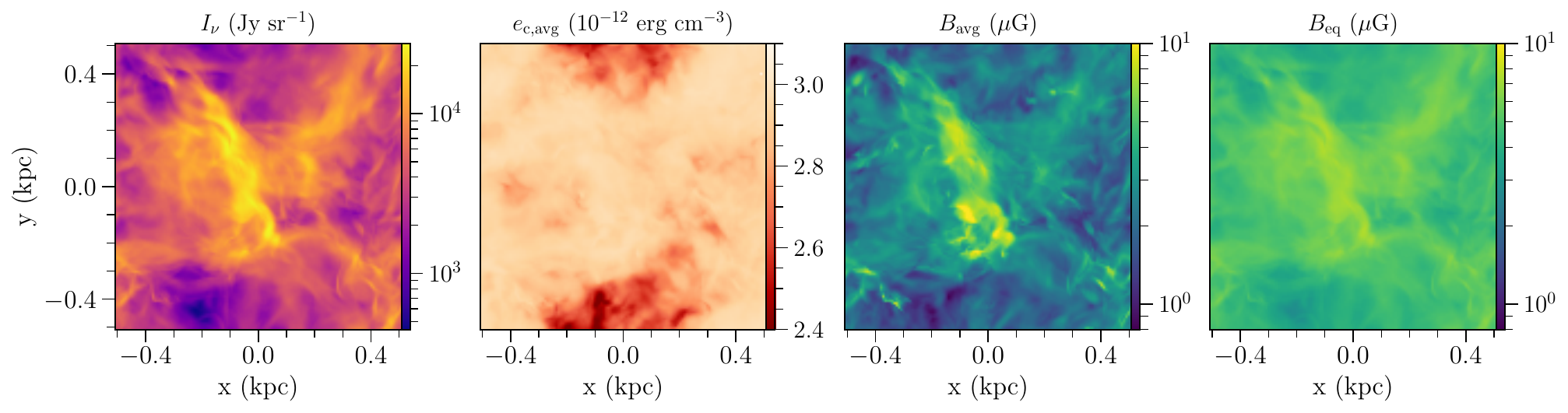}
    \caption{Vertically integrated and mass-averaged quantities from one snapshot of the {\tt R8} simulation. From left to right, the four panels show synchrotron intensity ($I_\nu$) at 1.5 GHz (\autoref{eq:Iv_synch}), the vertical mass-weighted average of total CR energy density ($e_\textrm{c,avg}$), the vertical mass-weighted average magnetic field strength ($B_{\rm avg}$), and the traditional equipartition estimate of the magnetic field strength ($B_{\rm eq}$) as determined by \autoref{eq:equipartition} from the synchrotron intensity. We can see that $e_\textrm{c,avg}$ is nearly constant spatially while $B_{\rm avg}$ has much greater variation. Therefore, the spatial distribution of $I_\nu$ primarily traces $B_{\rm avg}$, and  the traditional equipartition estimate fails at small scales.}
    \label{fig:snapshot}
\end{figure*}

In \autoref{fig:snapshot} we present the face-on view of one {\tt R8} snapshot including $I_\nu$ at 1.5 GHz as well as the vertically mass-weighted average values of the total CR energy density ($e_\textrm{c,avg}$) and simulated magnetic field strength ($B_{\rm avg}$). We also include the magnetic field strength determined by the equipartition estimate, $B_{\rm eq}$.

Qualitatively, the maps of $I_\nu$, $B_{\rm avg}$, and $B_{\rm eq}$ have similar spatial structure. However, $B_{\rm avg}$ has much larger variation in magnitude than $B_{\rm eq}$. Compared to the other three panels, $e_\textrm{c,avg}$ is smoother and has very little variation in magnitude across the simulation box. From only these snapshots, we can anticipate that equipartition between the magnetic field and CRs will not necessarily hold on the simulation-resolution scale.

\begin{figure}
    \centering
    \includegraphics[width=\linewidth]{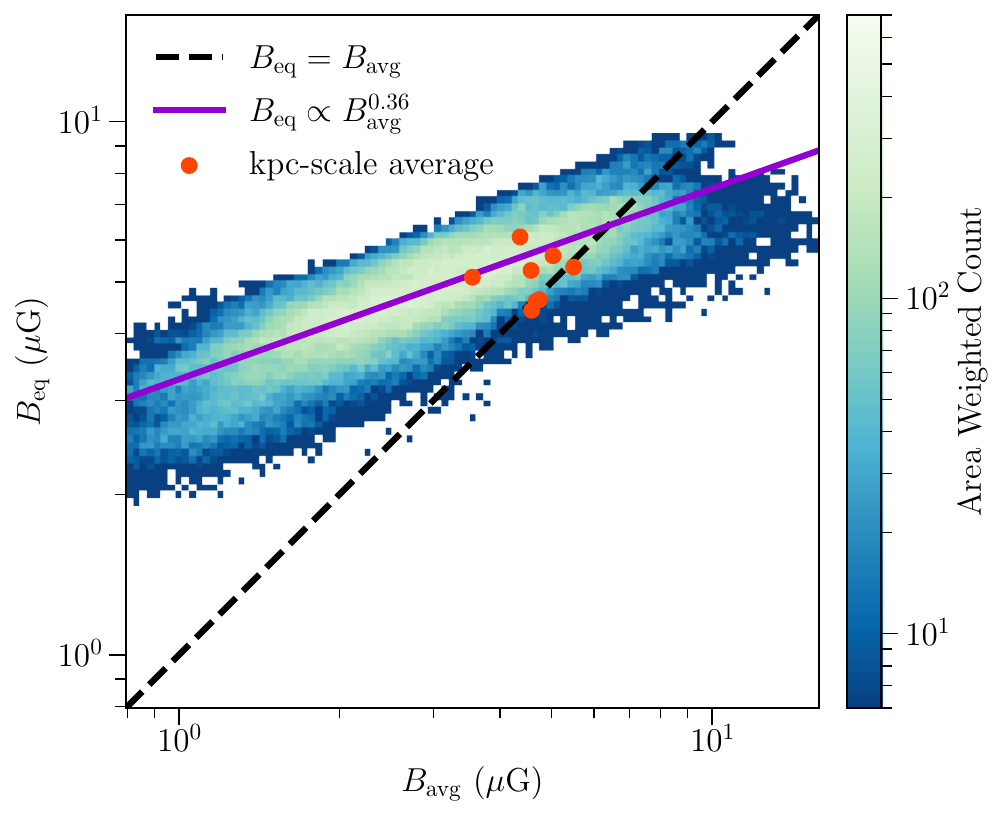}
    \caption{The blue distribution represents a two-dimensional histogram of the traditional equipartition magnetic field strength, $B_{\rm eq}$ (\autoref{eq:equipartition}), compared to the measured (mass-weighted) average magnetic field strength, $B_{\rm avg}$. The distribution includes each pixel in projected maps across all {\tt R8} model snapshots. The best fit power law to this distribution is included as a purple line with $B_{\rm eq} \propto B_{\rm avg}^{0.36}$. The black, dashed line represents $B_{\rm eq} = B_{\rm avg}$. The red points represent the values of $B_{\rm eq}$ and $B_{\rm avg}$ averaged over each kpc-scale simulation snapshot individually. While the large-scale snapshot-averaged value  of $B_{\rm eq}$ estimates the large-scale average of $B_{\rm avg}$ well, the pixel-by-pixel comparison shows that $B_{\rm eq}$ fails to recover the true $B_{\rm avg}$ at smaller scales.}
    \label{fig:eq_dist}
\end{figure}

In \autoref{fig:eq_dist}, the two-dimensional histogram compares the values of $B_{\rm eq}$ and $B_{\rm avg}$ from every pixel across all {\tt R8} snapshots. We also include the kpc-scale mean value of both $B_{\rm eq}$ and $B_{\rm avg}$ by averaging across each snapshot individually. The kpc-averages of  $B_{\rm eq}$ and $B_{\rm avg}$ are consistent with each other (within a factor of 1.4) for each snapshot, but the distribution showing values from individual (projected) pixels does not have the same agreement. We include a linear best-fit relating log$(B_{\rm eq})$ and log$(B_{\rm avg})$ in purple. If $B_{\rm eq}$ were a good estimator of $B_{\rm avg}$ we would expect this relationship to have unity slope, matching the black, dashed line. Instead, we find that $B_{\rm eq} \propto B_{\rm avg}^{0.36}$. 

We can understand this power-law scaling between $B_{\rm eq}$ and $B_{\rm avg}$ by considering the derivation of the equipartition magnetic field estimate. From \autoref{eq:Iv}, we know $I_\nu \propto B^{\alpha + 1}n_\mathrm{e,0}$, with $n_\mathrm{e,0} \propto e_c$. Based on the smooth CR distribution seen in \autoref{fig:snapshot}, $n_{\rm e,0}$ is essentially constant, independent of $B$. Substituting $I_\nu \propto B^{\alpha + 1}$ into \autoref{eq:equipartition}, we see that $B_{\rm eq} \propto B^{(\alpha + 1) / (\alpha + 3)}=B^{(\gamma + 1) / (\gamma + 5)}$. The CRE spectral index, $\gamma_e$, is both measured in our simulations \citep{Linzer2025} and observationally estimated to be between 2 and 3.3 for CREs between $\sim1-10$ GeV which contribute primarily to 1.5-3 GHz emission in environments with $B\sim 1-10\;\mu$G \citep[e.g.][]{padovani_spectral_2021}. This corresponds to $\alpha$ between 0.5 and 1.15. Therefore, we may expect $B_{\rm eq} \propto B^{0.4-0.5}$ when using \autoref{eq:equipartition}.

If we decrease the simulated image resolution by averaging over multiple pixels\footnote{To reduce the image resolution by a factor of two, for example, we average $I_\nu$ over squares of two by two pixels. We then find the magnetic field at lower resolution by averaging the magnitude of the magnetic field strength in cubes of eight cells. We find similar results whether we find the volume- or mass-weighted average of magnetic field strength.}, we find consistent results: the slope between $B_{\rm eq}$ and $B_{\rm avg}$ remains approximately constant, meaning that the purple line stays nearly unchanged in \autoref{fig:eq_dist}. At the same time, the range of magnetic field values decreases with resolution and eventually coincides with the red data points when averaging over the full simulation box. 

\subsubsection{{\tt R8 Arm} Model}

\begin{figure*}
    \centering
    \includegraphics[width=\linewidth]{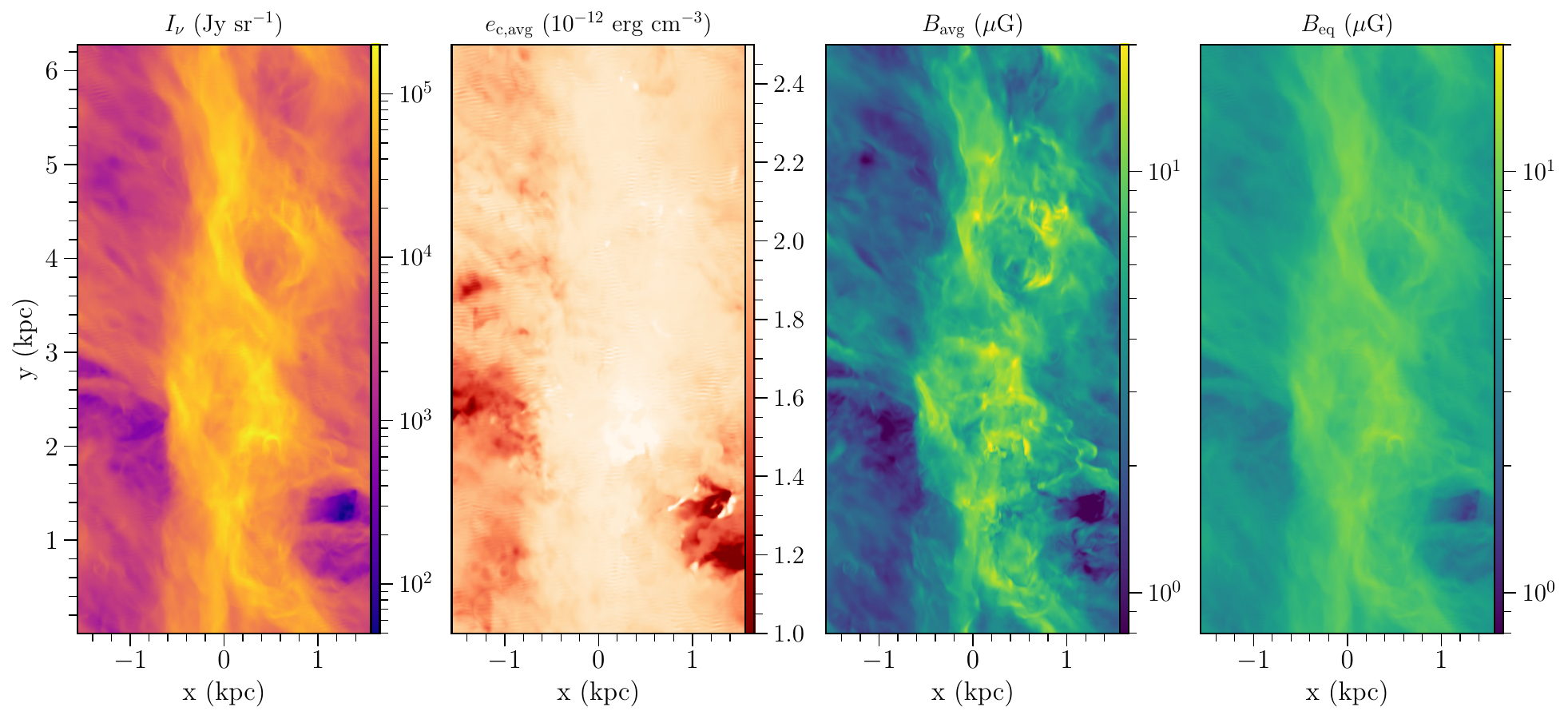}
    \caption{The same as in \autoref{fig:snapshot} for a snapshot of the {\tt R8 Arm} simulation. As in \autoref{fig:snapshot}, we see that $e_{c, \rm avg}$ is approximately constant while $B_{\rm avg}$ has greater variation. The spatial variation of $I_\nu$ is again determined by $B_{\rm avg}$, and $B_{\rm eq}$ does not estimate the small-scale magnetic field well.}
    \label{fig:arm_snap}
\end{figure*}

To supplement the {\tt R8} models originally described in \cite{Linzer2025}, we also post process and analyze snapshots from the {\tt R8 Arm} model. These additional models allow us to evaluate the traditional equipartition estimate across a wider range of spatial scales and physical conditions. In \autoref{fig:arm_snap} we present one snapshot of this model as in \autoref{fig:snapshot}. The {\tt R8 Arm} model, like the {\tt R8} simulation, shows a very smooth vertically-averaged CR energy distribution compared to the magnetic field strength. The arm structure is visible in all panels but is most prominent in $I_\nu$ and in the magnetic fields. While the spatial structure is similar in all panels, $B_{\rm eq}$ has less variation when compared to the true magnetic field strength.

\begin{figure}
    \centering
    \includegraphics[width=\linewidth]{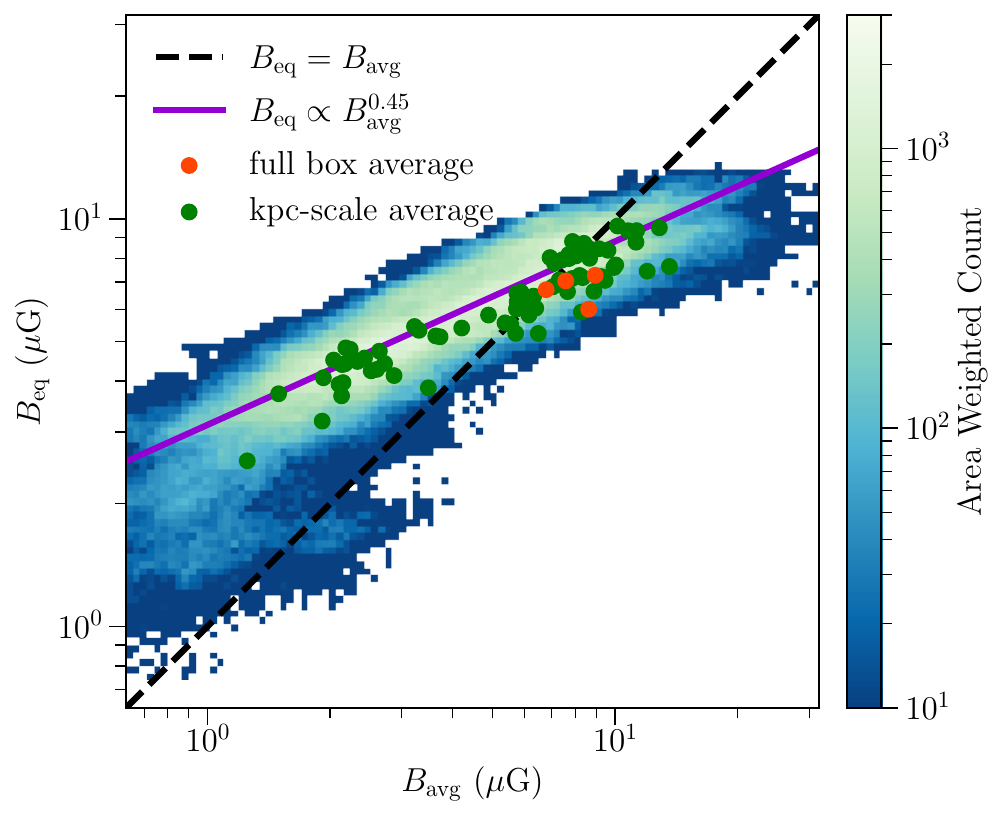}
    \caption{As in \autoref{fig:eq_dist}, now considering the {\tt R8 Arm} model. The blue distribution represents a two-dimensional histogram of the equipartition magnetic field strength, $B_{\rm eq}$, compared to the simulated, mass-averaged magnetic field strength, $B_{\rm avg}$. The best fit power law to this distribution is included as a purple line with $B_{\rm eq} \propto B_{\rm avg}^{0.45}$. The black, dashed line represents $B_{\rm eq} = B_{\rm avg}$. The red points represent the values of $B_{\rm eq}$ and $B_{\rm avg}$ averaged over the full simulation snapshots. The green points represent the averages over non-overlapping, 1 kpc-scale patches of each snapshot. 
    }
    \label{fig:arm_dist}
\end{figure}

As in \autoref{sec:R8_eq}, we estimate the magnetic field strength using the traditional equipartition estimate of \autoref{eq:equipartition}. We compare the values of $B_{\rm eq}$ to $B_{\rm avg}$ in each pixel in \autoref{fig:arm_dist}. Similar to the results for the {\tt R8} model, we see that $B_{\rm eq}$ does not agree with the true magnetic field at the simulation resolution ($\sim$12 pc). The best power-law fit of $B_{\rm eq}$ as a function of $B_{\rm avg}$ gives $B_{\rm eq} \propto B_{\rm avg}^{0.45}$. The reason for this sublinear scaling is the same as discussed in \autoref{sec:R8_eq}.

Because the {\tt R8 Arm} simulation box is larger than the {\tt R8} model, we can compare the magnetic field strength to the equipartition estimate averaged over non-overlapping kpc-scale patches. We include these results in \autoref{fig:arm_dist} as green points. For the original, kpc-scale {\tt R8} models, we found that the average equipartition magnetic field over the entire (1 kpc$^2$) simulation box was a good estimate of the true magnetic field (see the red points in \autoref{fig:eq_dist}). In the {\tt R8 Arm} model, however, we do not see good agreement between $B_{\rm eq}$ and $B_{\rm avg}$ at the kpc scale. Instead, the green points follow roughly the same scaling between $B_{\rm eq}$ and $B_{\rm avg}$ as the values at the simulation resolution. 

If we average over the full {\tt R8 Arm} snapshots, shown in the red points of \autoref{fig:arm_dist}, we do see approximate agreement between $B_{\rm avg}$ and $B_{\rm eq}$, within a factor 1.4. Thus, for both the {\tt R8} and {\tt R8 Arm} simulations, we find approximate agreement with the equipartition assumption to hold over the full simulation box, independent of the box size. Equipartition at the kpc scale, however, does not in general hold in the larger simulations. The widths of both the arm and inter-arm regions in the {\tt R8 Arm} model are $\sim1$ kpc. While the CR energy density varies by a factor of $\sim 2$ between the two regions, the magnetic field strength varies by more than an order of magnitude. Therefore, equipartition cannot always hold at the $\sim1$ kpc when considering both of these regions.

\subsection{Alternative methods to estimate the magnetic field strength}
\label{sec:alt}

In both the {\tt R8} and {\tt R8 Arm} models, we find that the CR and magnetic energy densities are in approximate equipartition when averaged across the entire simulation box (\autoref{sec:measured_e}), and the traditional approach using synchrotron emission reflects this (\autoref{sec:eq_est}). At smaller scales, however, we find that the equipartition estimate given by \autoref{eq:equipartition} does not accurately predict the true magnetic field strength. Therefore, we will consider alternative methods to estimate the magnetic field strength from synchrotron emission including estimates which rely on additional observations. The primary difference in each of these approaches compared to the traditional method will be to estimate the CR energy density $e_\mathrm{c}$ as a constant value averaged over large scales, $\langle e_\mathrm{c}\rangle$. 

\subsubsection{Predicting local $B$ for known $\langle e_\mathrm{c}\rangle$}
\label{sec:known_ec}

We can first test how well we can recover the true magnetic field strength if we use the simulated values of $e_{\rm c}$. To do so, we do not make the assumption that $e_{\rm c} = B^2/8\pi$ in \autoref{eq:eq_ec}. Instead we let $e_{\rm c} = \langle e_{\rm c}\rangle$ where $\langle e_{\rm c}\rangle$ is the volume-averaged CR energy density within $|z| < 300$ pc in each simulation snapshot. Then, rather than using \autoref{eq:equipartition} to predict the magnetic field strength, we instead find from \autoref{eq:eq_ec} and \autoref{eq:B_perp}
\begin{multline}
    B_{\rm pred, \langle e_{\rm c}\rangle} = \\ \left[ \frac{(2\alpha + 1) (K_0 + 1) E_{\rm p}^{1-2\alpha}(\nu/2c_1)^{\alpha}}{2(2\alpha - 1)c_2(\alpha)  c_4(i) } \frac{I_\nu}{\ell_z \langle e_{\rm c}\rangle}\right]^\frac{1}{\alpha + 1}.
    \label{eq:Beq_ec}
\end{multline}

\begin{figure}
    \centering
    \includegraphics[width=\linewidth]{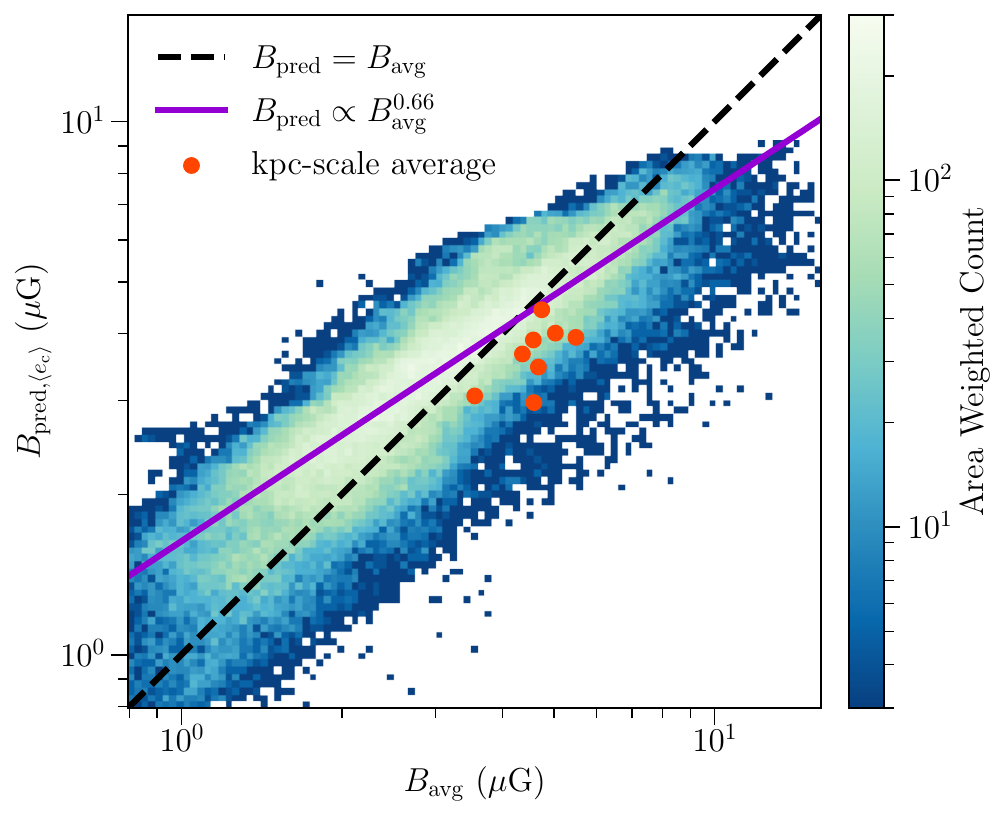}
    \caption{The blue distribution represents a two-dimensional histogram of the predicted magnetic field strength, $B_{\textrm{pred,} \langle e_{\rm c}\rangle}$ (\autoref{eq:Beq_ec}), compared to the simulated, mass-averaged magnetic field strength, $B_{\rm avg}$ across all {\tt R8} snapshots. The red points represent the values of $B_{\textrm{pred,} \langle e_{\rm c}\rangle}$ and $B_{\rm avg}$ averaged over each kpc-scale simulation snapshot individually. The black, dashed line represents $B_{\rm pred} = B_{\rm avg}$. The best fit power law to this distribution is included as a purple line given by $B_{\rm pred} \propto B_{\rm avg}^{0.66}$. This improves over the traditional equipartition estimate at small scales (\autoref{fig:eq_dist}).}
    \label{fig:eq_ec}
\end{figure}

In \autoref{fig:eq_ec}, we show the two-dimensional histogram comparing the true magnetic field strength to the value predicted by \autoref{eq:Beq_ec} in each pixel across all {\tt R8} snapshots. We find that this alternative estimate improves over the traditional equipartition estimate, with $B_{\rm pred, \langle e_{\rm c}\rangle} \propto B_{\rm avg}^{0.66}$ compared to $B_{\rm eq} \propto B_{\rm avg}^{0.36}$ (\autoref{fig:eq_dist}). On average, $B_{\rm pred, \langle e_{\rm c}\rangle}$ overestimates the true magnetic field strength by approximately $10\%$ when comparing the two values pixel by pixel. Comparing the mass-averaged values across the simulation box (red points in \autoref{fig:eq_ec}), $B_{\rm pred, \langle e_{\rm c}\rangle}/B_\mathrm{avg}$ varies over a range $0.65-0.94$. 

Even when using the true CR energy density, there is still disagreement between $B_{\rm pred, \langle e_{\rm c}\rangle}$ and $B_{\rm avg}$ in both the scaling between the two values as well as the magnitude. The fact that we find $B_{\rm pred, \langle e_{\rm c}\rangle} \propto B_{\rm avg}^{0.66}$ rather than $B_{\rm pred, \langle e_{\rm c}\rangle} \propto B_{\rm avg}$ can be explained in part by the implicit assumptions in \autoref{eq:Iv} when estimating $I_\nu$. In particular, the synchrotron emission is assumed to be produced by a slab of height $\ell_z$ with constant $n_e$, $\gamma_e$, and $B_\perp$. However, these values all vary as a function of height above the galactic midplane, and do not change in the same way.

Furthermore, while the standard choices of $\ell_z = 1$ kpc and $K_0 = 100$ are roughly consistent with our simulations of CR transport in the {\tt R8} environment, more generally they would vary with local gas and CR properties. By choosing a different value of $\ell_z$ (or $K_0$), we could change the overall magnitude of $B_{\rm pred, \langle e_{\rm c}\rangle}$ to more closely match $B_{\rm avg}$. In our simulations, the actual box-averaged ratio of $K_0=n_\mathrm{c,p}/n_\mathrm{c,e}$ is approximately 80 between CR energies of $\sim1-10$ GeV. If we were to use this value, $\ell_z = 0.7$ kpc results in the best agreement between $B_{\rm pred, \langle e_{\rm c}\rangle}$ and $B_{\rm avg}$ when averaged over the box. This ``best-fit'' value of $\ell_z$ could change for many reasons, however. For example, if we defined $B_{\rm avg}$ using an emission- or volume-weighted average rather than mass-weighted, the true magnetic field would be different necessitating a different normalization of $B_{\rm pred}$. Also, changing the observed synchrotron frequency results in a slightly different value of $B_{\rm pred}$.

\cite{2024MNRAS.52711707P} use galaxy-scale simulations of CR transport to test the validity of the equipartition estimate. They find that their predicted magnetic field strength underestimates the true value, and renormalize their predicted values through the introduction of a volume-filling factor, $f_V$. The physical interpretation of $f_V$ is to account for the fact that the majority of synchrotron emission is produced in gas which makes up only a small fraction of the total volume, and this fraction varies as a function of galactic radius. Numerically, this is equivalent to a value of $\ell_z$ that changes based on the ISM environment, as $f_V$ appears through multiplying $\ell_z$ in \autoref{eq:equipartition}.

\subsubsection{Setting $\langle e_\mathrm{c}\rangle$ from large-scale equipartition}
\label{sec:large-scale-eq}

While \autoref{eq:Beq_ec} improves upon the traditional equipartition estimate on small scales, $\langle e_{\rm c} \rangle$ is not known in reality. To apply this method to possible observations, we consider alternative estimates of $\langle e_{\rm c}\rangle$. In the first alternative approach, we make use of the fact that we find equipartition to hold approximately at large scales. Therefore, we consider setting $\langle e_\mathrm{c} \rangle = \langle B^2 \rangle/8\pi$ where $\langle B^2 \rangle$ is the volume-averaged values of $B^2$ determined as follows. 

Using \autoref{eq:eq_ec} with \autoref{eq:B_perp} and $\gamma = 2\alpha + 1$ we have
\begin{equation}
    \label{eq:ec_A}
    e_c B^{\alpha + 1} = A I_{\nu}
\end{equation}
where
\begin{equation}
    A = \frac{(2\alpha + 1) (K_0 + 1) E_{\rm p}^{1-2\alpha}(\nu/2c_1)^{\alpha}}{2(2\alpha - 1)c_2(\alpha)  c_4(i) \ell_z}.
\end{equation}
By taking both sides of \autoref{eq:ec_A} to the power $2/(\alpha +1)$ and averaging, we find
\begin{equation}
    \langle e_c^{\frac{2}{\alpha + 1}} B^{2}\rangle = \langle A^{\frac{2}{\alpha + 1}} I_{\nu}^{\frac{2}{\alpha + 1}} \rangle.
\end{equation}

We now make the assumption that $e_c$ is approximately constant spatially, $e_\mathrm{c}\approx \langle e_\mathrm{c}\rangle $. Additionally, we assume that the CR spectral slope is similar in each pixel across the simulation box and we can therefore use a single, average value of $\bar{\alpha}$ defined as
\begin{equation}
    \bar{\alpha} = -\frac{{\rm log}(\langle I_{\nu_1}\rangle /\langle I_{\nu_2}\rangle )}{{\rm log}(\nu_1/\nu_2)}\,
\end{equation}
where $\langle I_{\nu}\rangle$ is the spatial average across the simulation snapshot. Therefore, we can write
\begin{equation}
    \langle e_c\rangle^{\frac{2}{\bar{\alpha} + 1}} \langle B^{2}\rangle = 8\pi\langle e_c\rangle^{\frac{\bar{\alpha} + 3}{\bar{\alpha} + 1}} = A^{\frac{2}{\bar{\alpha} + 1}} \langle I_{\nu}^{\frac{2}{\bar{\alpha} + 1}} \rangle.
\end{equation}
Thus,
\begin{equation}\label{eq:ecavg}
    \langle e_c\rangle = \frac{A^\frac{1}{\bar{\alpha} + 3} \langle I_{\nu}^{\frac{2}{\bar{\alpha} + 1}}\rangle^\frac{\bar{\alpha} + 1}{\bar{\alpha} + 3}}{(8 \pi)^\frac{\bar{\alpha} + 1}{\bar{\alpha} + 3}}.
\end{equation}

Returning to \autoref{eq:ec_A}, we can now solve for the local value of $B$ using the local value of $I_\nu$ and using \autoref{eq:ecavg} for $e_\mathrm{c}$, taking $\alpha \rightarrow\bar{\alpha}$. This leads to the final expression:
\begin{multline} 
\label{eq:Beq_avg}
 B_{\mathrm{pred}, \langle B^2 \rangle} = I_\nu^{\frac{1}{\bar{\alpha} + 1}} \times \\\left[ \frac{4 \pi (2\bar{\alpha} + 1) (K_0 + 1) E_{\rm p}^{1-2\bar{\alpha}}(\nu/2c_1)^{\bar{\alpha}}}{(2\bar{\alpha} - 1)c_2(\bar{\alpha})  c_4(i) \ell_z 
 \langle I_\nu^{\frac{2}{\bar{\alpha} + 1}} \rangle
}
 \right]^\frac{1}{\bar{\alpha} + 3}.
\end{multline}

\begin{figure*}
    \centering
    \includegraphics[width=\linewidth]{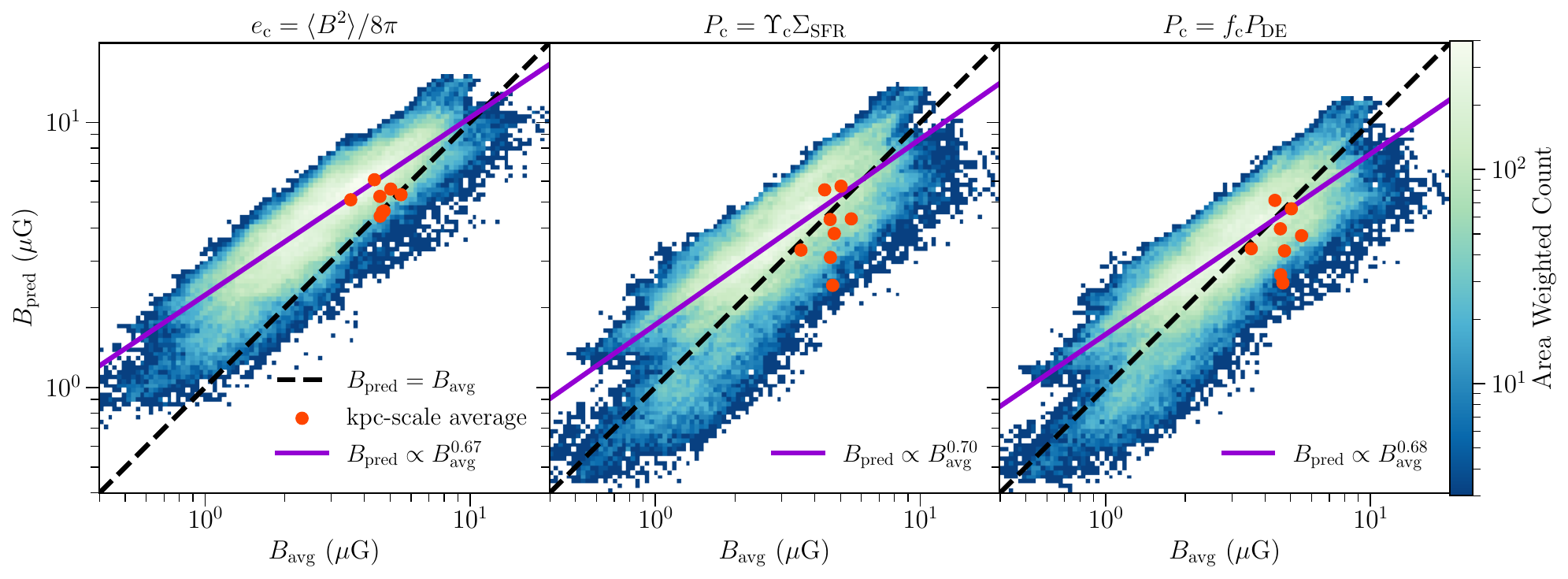}
    \caption{Each panel shows a two-dimensional histogram comparing one prediction of the magnetic field strength to the simulated, mass-averaged value. These distributions include all pixels across all {\tt R8} snapshots. The left panel estimates the magnetic field using \autoref{eq:Beq_avg}, where the total CR energy density $e_\mathrm{c}$ is assumed to be a spatially constant value determined by equipartition with the spatial average value of the magnetic energy density. The middle panel also assumes spatially constant $e_\mathrm{c}$, adopting a calibrated relationship between cosmic ray pressure and star formation rate per unit area, $\Sigma_{\rm SFR}$. The right panel assumes a spatially constant  $e_\mathrm{c}$ and estimates the mean CR pressure as a fraction of the total dynamical equilibrium pressure (or ISM weight) (\autoref{eq:Beq_W}). In all three panels, the black, dashed line shows $B_{\rm pred} = B_{\rm avg}$ and the purple line shows the best-fit power law to the distribution. The red points represent the box-averaged values of $B_{\rm pred}$ and $B_{\rm avg}$ from each simulation snapshot. In all three models we see an improvement in the scaling of $B_{\rm pred}$ relative to $B_{\rm avg}$ compared to the distribution shown in \autoref{fig:eq_dist}.}
    \label{fig:eq_avg}
\end{figure*}

We compare this new prediction to the simulated magnetic field in the left panel of \autoref{fig:eq_avg}. As in \autoref{fig:eq_dist} and \autoref{fig:eq_ec}, we show the comparison between the observationally predicted and simulated magnetic field strength in each pixel across all snapshots of the {\tt R8} model. The best fit power law to this distribution is $B_{\mathrm{pred}} \propto B_{\rm avg}^{0.67}$, the same as in \autoref{fig:eq_ec} when using the simulated value of $\langle e_{\rm c} \rangle$. The box-averaged value of $B_{\mathrm{pred}, \langle B^2 \rangle}$, shown in the red points, differs from the true magnetic field strength by a factor of $B_{\mathrm{pred}, \langle B^2 \rangle} / B_{\rm avg}\sim1-1.4$.

\subsubsection{Scaling $\langle e_\mathrm{c}\rangle$ with SFR}
\label{sec:SFR_scaling}

The estimate in \autoref{eq:Beq_avg} still relies on the assumption of equipartition between the CR and magnetic energy density, which is approximately true for the present models if we are only considering large scales \autoref{sec:measured_e}, but need not be true in all environments. A different method would be to estimate $\langle e_\mathrm{c}\rangle$ from independent observable quantities. This alternative approach employs physical  relationships between $\langle e_\mathrm{c}\rangle$ and other galactic observables that are calibrated with numerical simulations. One alternative of this kind would employ observations that provide an estimate of the star formation rate per unit area, $\Sigma_{\rm SFR}$, which is proportional to the energy injection rate from supernovae provided core-collapse systems dominate. 

\begin{comment}
From Equation 28 of \citet{Armillotta2025}, the midplane pressure of 1 GeV CRs is given by
\begin{multline}
\label{eq:P_c}
P_\mathrm{c} ({\rm 1 \; GeV}) = \\ \dfrac{F_\mathrm{in} ({\rm 1 \; GeV})}{4 v_\mathrm{c,eff}({\rm 1 \; GeV})  + 3 \Lambda_\mathrm{coll}({\rm 1 \; GeV}) \Sigma_\mathrm{gas}/(2 \mu_\mathrm{H} m_\mathrm{p})}. 
\end{multline}
We note that here we consider the pressure of only the 1 GeV CR protons rather than the total CR population as in \citet{Armillotta2025}.   

$F_\mathrm{in}$ represents the CR flux injected vertically to each half of the disk and is given by
\begin{equation}\label{eq:Fin_def}
F_\mathrm{in} = 0.5 \epsilon_\mathrm{c} ({\rm 1 \; GeV})  E_\mathrm{SN} \Sigma_\mathrm{SFR}/m_\star
\end{equation}
where $E_\mathrm{SN} = 10^{51}$~erg is the energy released from a single SN event, $\epsilon_\mathrm{c} ({\rm 1 \; GeV})$ is the assumed fraction of SN energy converted into acceleration of 1 GeV CRs,
\end{comment}
From Equation 28 of \citet{Armillotta2025}, the midplane pressure of CRs in a given energy bin is
\begin{equation}
\label{eq:P_c}
P_\mathrm{c} = \dfrac{F_\mathrm{in}}{4 v_\mathrm{c,eff} + 3 \Lambda_\mathrm{coll} \Sigma_\mathrm{gas}/(2 \mu_\mathrm{H} m_\mathrm{p})}. 
\end{equation}
Here, $F_\mathrm{in}$ represents the CR flux injected vertically to each half of the disk and is given by
\begin{equation}\label{eq:Fin_def}
F_\mathrm{in} = 0.5 \epsilon_\mathrm{c}  E_\mathrm{SN} \Sigma_\mathrm{SFR}/m_\star
\end{equation}
where $E_\mathrm{SN} = 10^{51}$~erg is the energy released from a single SN event, $\epsilon_\mathrm{c}$ is the fraction of SN energy transferred to CRs in the specified energy bin, and $m_\star$ is the mass of stars per SN ($95.5 \, M_\odot$ in \citealt{kim_three-phase_2017} from a Kroupa IMF). CR energetic losses are included through $\Lambda_\mathrm{coll}$ where the CR energy loss rate\footnote{The loss term is quite small compared to the transport term for CR protons with $E_\mathrm{k} \geq 1$~GeV in the environment considered here, but can become more important in high-density environments.} is given by $dE/dt = \Lambda_\mathrm{coll} n_H e_c$, with $\Lambda_\mathrm{coll}$ being energy dependent.

The term $v_\mathrm{c,eff}$, is the effective CR propagation speed out of the midplane region. Note that $v_\mathrm{c,eff}$ varies as a function of CR energy because it includes diffusive as well as advection and streaming components. At low energy, advection and streaming are expected to dominate, while at higher CR energy, diffusion becomes more important \citep[see Section 4 of][for a 1D model and comparison to simulations]{Armillotta2025}. 

\citet{2025arXiv250903519H} used a set of post-processed TIGRESS simulations modeling CR protons with kinetic energy $E_{\mathrm{k}} \approx 1 $ GeV in multiple different environments to calibrate $v_\mathrm{c,eff}$ as a function of $\Sigma_{\rm SFR}$. These environments include the {\tt R8}, {\tt R8 Arm}, {\tt R4}, and {\tt R2} models from \citet{kim_first_2020, kim_arm_2020}, which have $\Sigma_{\rm gas} \sim 10-70 \;\textrm{M}_{\odot} \textrm{ pc}^{-2}$ and $\Sigma_{\rm SFR} \sim 5\times10^{-3}-1 \;\textrm{M}_{\odot} \textrm{ kpc}^{-2} \textrm{ yr}^{-1}$. The calibration reported in \citet[][see their Eq. 33]{2025arXiv250903519H} is
\begin{multline}
\label{eq:veff_fit}
 v_\mathrm{c,eff} ({E_\mathrm{k} \approx \rm 1 \; GeV}) =\\  21\ \mathrm{km\ s^{-1} }\left(\frac{\Sigma_\mathrm{SFR}}{0.01~\mathrm{M_\odot~ pc^{-2}~Myr^{-1}}}\right)^{0.2}. 
\end{multline}

Substituting \autoref{eq:Fin_def} and  \autoref{eq:veff_fit} in \autoref{eq:P_c}, and assuming a value for $\epsilon_\mathrm{c}$, we can obtain 
\begin{equation}
\label{eq:P_ups}
P_{\rm c} ({E_\mathrm{k} \approx \rm 1 \; GeV}) = \Upsilon_{\rm c} ({E_\mathrm{k} \approx \rm 1 \; GeV}) \Sigma_{\rm SFR} 
\end{equation}
where $\Upsilon_{\rm c} ({E_\mathrm{k} \approx \rm 1 \; GeV})$ is a theoretically-predicted CR feedback ``yield'' for 1 GeV CRs. However, we are interested in the total CR energy density, not just that of the 1 GeV CRs. Therefore, we will assume a CR spectrum consistent with direct observations with CR number density per unit energy given by $n(E) \propto E^{-2.7}$ \citep[e.g.][]{grenier_nine_2015}; this spectrum is also consistent with our simulation results reported in \citet{Armillotta2025}. Using \autoref{eq:P_ups} to find the CR pressure at 1 GeV, and integrating the CR distribution function over all energies $> 1$ GeV, we then find an integrated CR yield given by
\begin{equation}
\label{eq:Upsilon_c}
   \Upsilon_{\rm c, int} = 132 \textrm{ km s}^{-1} \left(\frac{\Sigma_{\rm SFR}}{0.01 \textrm{ M}_\odot \textrm{ kpc}^{-2}\textrm{ yr}^{-1}}\right)^{-0.23}.
\end{equation}

The value of $\Upsilon_{\rm c, int}$ reported here is approximately 50\% smaller compared to the value of $\Upsilon_{\rm c}$ in \citet{2025arXiv250903519H}. This is due to the fact that they model only a single energy bin with scattering rate appropriate for CR protons at 1 GeV. Our work here includes spectrally resolved CRs up to 100 GeV. The higher energy CRs have higher diffusion coefficient (lower scattering rate) and therefore more efficiently escape the disk, which corresponds to larger $v_\mathrm{c,eff}$ at higher CR energy compared to $v_\mathrm{c,eff}({E_\mathrm{k} \approx \rm 1 \; GeV})$ (see Eq. 21 of \citealt{Armillotta2025}). Since $\Upsilon_\mathrm{c}$ varies inversely with $v_\mathrm{c,eff}$, the value of $\Upsilon_\mathrm{c,int}$ must be smaller than the value in \citet{2025arXiv250903519H}.

The CR energy density can then be written as $e_{\rm c} = 3 \Upsilon_{\rm c, int} \Sigma_{\rm SFR}$ and the predicted magnetic field strength from \autoref{eq:ec_A} is 
\begin{multline}
    B_{\rm pred, SFR} = \\ \left[ \frac{(2\alpha + 1) (K_0 + 1) E_{\rm p}^{1-2\alpha}(\nu/2c_1)^{\alpha}}{6(2\alpha - 1)c_2(\alpha)  c_4(i) } \frac{I_\nu}{\ell_z \Upsilon_{\rm c, int} \Sigma_{\rm SFR}}\right]^\frac{1}{\alpha + 1}.
    \label{eq:Beq_SFR}
\end{multline}
Here, $\Sigma_{\rm SFR}$ is the average value over the simulation box; in observations, a similarly large-scale average would be used. This prediction for the magnetic field strength is shown in the middle panel of \autoref{fig:eq_avg}. The best fit power-law of this distribution gives $B_{\rm pred} \propto B_{\rm avg}^{0.70}$ and the box-averaged value of $B_{\rm pred}/B_{\rm avg}$ varies between 0.5-1.3. 

\subsubsection{Scaling $\langle e_\mathrm{c}\rangle$ with ISM weight}
\label{sec:weight_scaling}

Alternatively, we could also assume the midplane CR pressure is some fraction of the total ISM weight. A simple estimate of the weight in regions of galaxies where the gas disk is thinner than the stellar disk is 
\begin{equation}\label{eq:PDEdef}
    \mathcal{W} \approx P_\mathrm{DE}= \frac{\pi G \Sigma_{\rm gas}^2}{2} + \Sigma_{\rm gas}\sigma_{\rm eff}\sqrt{2 G \rho_{\rm sd}}.
\end{equation}
Here, we estimate the weight, $\mathcal{W}$, using a formula for dynamical equilibrium pressure, $P_\mathrm{DE}$, that has been adopted in a number of observational studies (e.g. \citealt{2020ApJ...892..148S} and references therein; see also  \citealt{2025ApJ...989...66V,2024ApJ...975..151H} for regime of applicability and more general forms). In this expression, $G$ is the gravitational constant and $\Sigma_{\rm gas}$ is the gas surface density, while the combined stellar and dark matter density is given by $\rho_{\rm sd}$, and $\sigma_{\rm eff}$ represents the effective vertical velocity dispersion. The form in \autoref{eq:PDEdef} has been demonstrated to agree well with the total measured midplane pressure in all thermal components, in a set of TIGRESS numerical simulations \citep{ostriker_pressure-regulated_2022}.

For our test, we take the values directly from the simulation. In the {\tt R8} model, $\rho_{\rm sd} = 0.0924\;M_\odot\textrm{ pc}^{-3}$. We estimate $\sigma_{\rm eff} = (\Sigma_{ijk}P_{\rm tot}/\Sigma_{ijk}\rho)^{1/2}$ where $\Sigma_{ijk}$ represents the sum over the whole simulation box. Here we take the total pressure as the sum of the thermal and effective turbulent pressure, $P_{\rm tot} = P_{\rm th} + P_{\rm turb}$. In reality, the total pressure includes an additional term, the Maxwell stress, representing the magnetic pressure and tension. However, as this value is not accessible observationally we include only $P_{\rm th}$ and $P_{\rm turb}$. The exclusion of the magnetic pressure term results in at most a 10\% reduction in $\sigma_{\rm eff}$ across the {\tt R8} snapshots, so does not have a significant effect on our final results.

\cite{ostriker_pressure-regulated_2022} show that, in the TIGRESS simulations, the weight is balanced by the sum of the midplane thermal, kinetic, and magnetic pressure, such that $\mathcal{W} \approx P_\mathrm{DE}\approx P_\mathrm{tot} = \Upsilon_{\rm tot} \Sigma_\mathrm{SFR}$, where $\Upsilon_{\rm tot}$ is the combined feedback yield from thermal, kinetic, and magnetic components.\footnote{Note that because the scale height of CRs is larger than that of the gas, the CR pressure would not be expected to contribute in balancing the ISM weight even if CRs had been included in the original TIGRESS simulation.}  We define the total midplane CR pressure as a fraction of the total weight 
\begin{equation}
P_{\rm c} = f_{\rm c} \mathcal{W}\approx f_\mathrm{c} P_\mathrm{DE}.  
\end{equation}
Since $P_{\rm c} = \Upsilon_{\rm c, int} \Sigma_{\rm SFR}$ and $\mathcal{W} = \Upsilon_{\rm tot} \Sigma_\mathrm{SFR}$, the expected fraction is $f_{\rm c} = \Upsilon_{\rm c, int}/\Upsilon_{\rm tot}$. \cite{ostriker_pressure-regulated_2022} fit $\Upsilon_{\rm tot}$ as a function of $P_\mathrm{DE}$ (see their Equation 25c), and find
\begin{equation}
    \Upsilon_{\rm tot} = 1030 \textrm{ km s}^{-1} \left(\dfrac{P_\mathrm{DE}/ k_B}{10^4 \textrm{ cm}^{-3} \textrm{ K}}\right)^{-0.21}.
\end{equation}

To find the relation between $\Upsilon_{\rm c, int}$ and $\mathcal{W}$, we again use the post-processed TIGRESS simulations from \citet{2025arXiv250903519H}, which provides the fit
\begin{equation}\label{eq:veff_PDE}
 v_\mathrm{c, eff} ({E_\mathrm{k} \approx \rm 1 \; GeV}) = 15\, \mathrm{km\ s^{-1} }\left(\frac{P_\mathrm{DE}/k_B}{10^4 \textrm{ cm}^{-3} \textrm{ K}}\right)^{0.26}.
\end{equation}
As in \autoref{sec:SFR_scaling}, we integrate over the CR spectrum and find the total yield to be
\begin{equation}\label{eq:Ups_PDE}
    \Upsilon_{\rm c, int} = 198 \textrm{ km s}^{-1} \left(\frac{P_\mathrm{DE}/k_B}{10^4 \textrm{ cm}^{-3} \textrm{ K}}\right)^{-0.29};
\end{equation}
note that the absolute value of the indices in \autoref{eq:veff_PDE} and \autoref{eq:Ups_PDE} differ slightly because of the loss term. Applying \autoref{eq:Ups_PDE} results in a value of
\begin{equation}
    f_{\rm c} = \frac{\Upsilon_{\rm c, int}}{\Upsilon_{\rm tot}} \approx 0.19 \left(\frac{P_\mathrm{DE}/k_B}{10^4 \textrm{ cm}^{-3} \textrm{ K}}\right)^{-0.08}.
\end{equation}

Substituting $e_{\rm c} = 3f_{\rm c}P_\mathrm{DE}$ into \autoref{eq:ec_A} results in a predicted magnetic field strength of
\begin{multline}
    B_{\rm pred, \mathcal{W}} = \\ \left[ \frac{(2\alpha + 1) (K_0 + 1) E_{\rm p}^{1-2\alpha}(\nu/2c_1)^{\alpha}}{6(2\alpha - 1)c_2(\alpha)  c_4(i) } \frac{I_\nu}{\ell_z f_{\rm c} P_\mathrm{DE}}\right]^\frac{1}{\alpha + 1},
    \label{eq:Beq_W}
\end{multline}
where for $P_\mathrm{DE}$ we use \autoref{eq:PDEdef} with $\sigma_\mathrm{eff}$ averaged across the simulation box. We evaluate \autoref{eq:Beq_W} for each of the {\tt R8} snapshots and plot the distribution in the right panel of \autoref{fig:eq_avg}. As with the alternate evaluations given by \autoref{eq:Beq_avg} and \autoref{eq:Beq_SFR}, we see a scaling $B_{\rm pred} \propto B_{\rm avg}^{0.68}$ that is closer to linear compared to the traditional equipartition approach. The box-averaged values of $B_{\rm pred}/B_{\rm avg}$ vary between 0.5-1.2.

The three alternative estimates show similar power-law relationships between $B_{\rm pred}$ and $B_{\rm avg}$ at the 8 pc scale, all of which are closer to linear than the unmodified equipartition estimate. Therefore, each of these new methods may provide a better estimate of the small-scale magnetic field strength, with caveats that we will discuss in \autoref{sec:disc_2}. The overall normalization of $B_{\rm pred}$, however, differs based on the estimator used for $e_c$. If we had good observational constraints on the magnetic field strength through alternative methods, for example in the Milky Way, we might constrain an overall renormalization factor.

\subsubsection{{\tt R8 Arm} Model}
\label{sec:arm_alt}

\begin{figure}
    \centering
    \includegraphics[width=\linewidth]{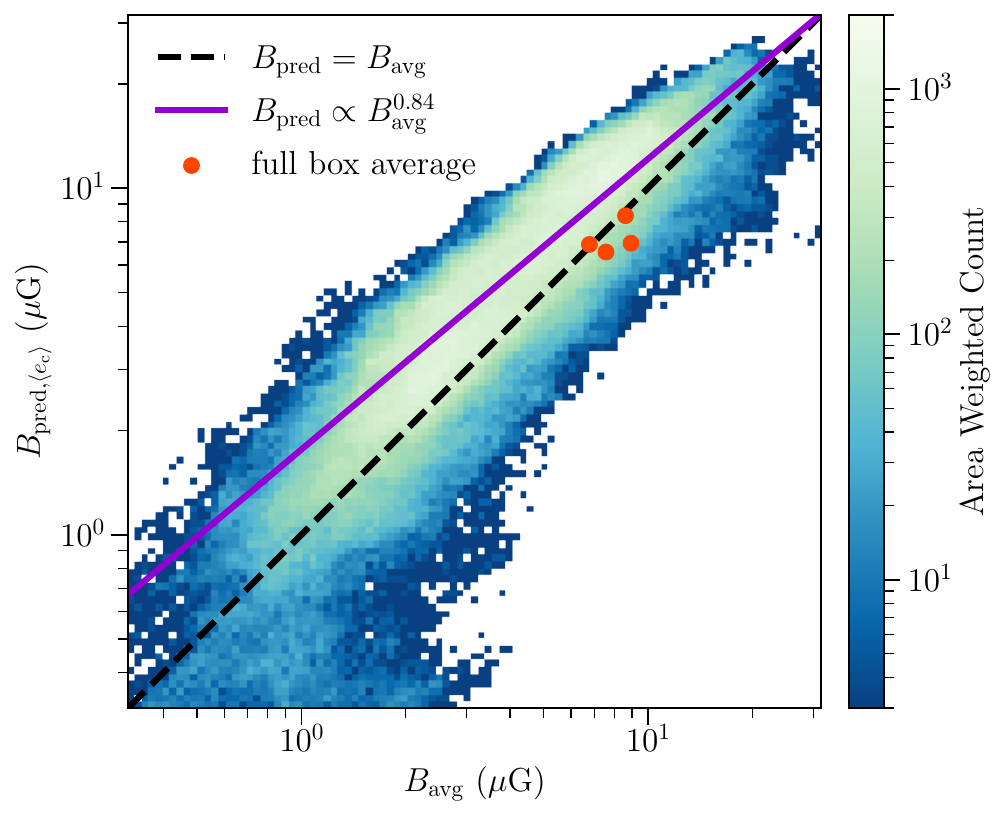}
    \caption{The same as in \autoref{fig:eq_ec} for the {\tt R8 Arm} model.}
    \label{fig:arm_ec}
\end{figure}

We also apply the alternative estimates to the {\tt R8 Arm} models. In \autoref{fig:arm_ec} we show the predicted magnetic field strength given by \autoref{eq:Beq_ec} using the simulated value of $\langle e_{\rm c} \rangle$.
As in the {\tt R8} model, this alternative estimate improves over the traditional approach at small scales. We find that $B_{\rm pred} \propto B_{\rm avg}^{0.84}$, which is significantly closer to a linear relationship between the true and predicted magnetic field strengths. 

\begin{figure*}
    \centering
    \includegraphics[width=\linewidth]{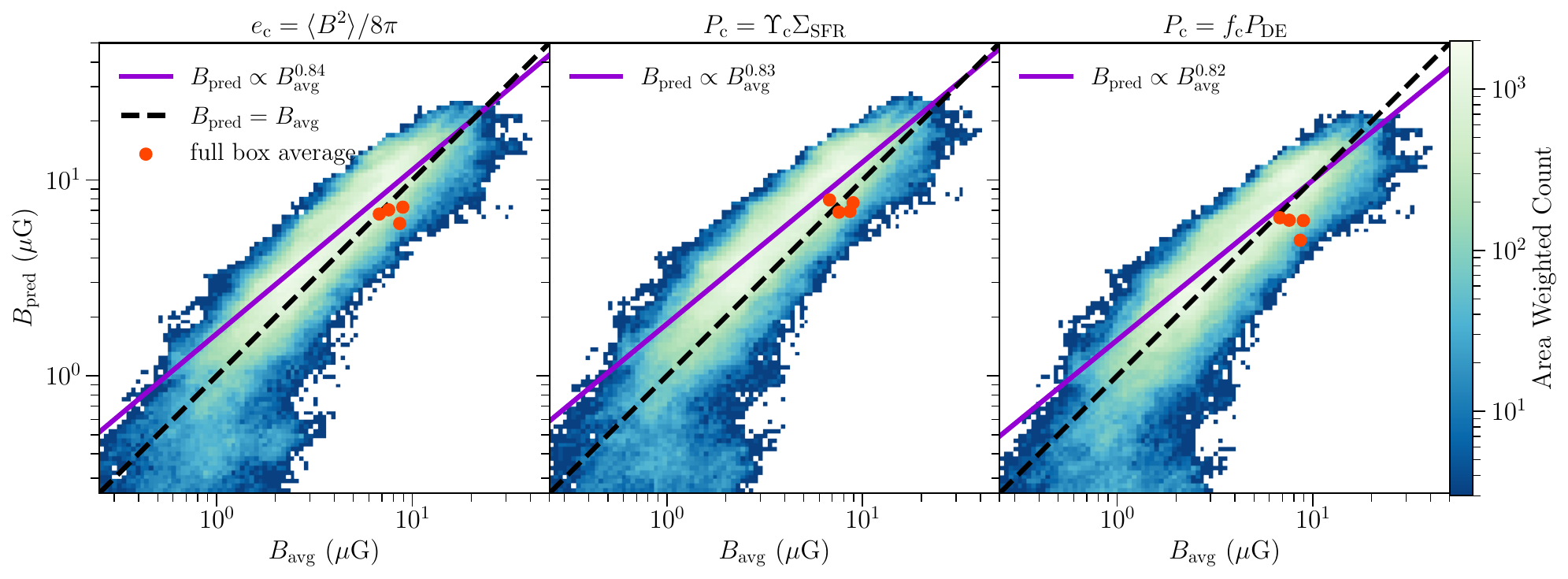}
    \caption{The same as in \autoref{fig:eq_avg} for the {\tt R8 Arm} model.}
    \label{fig:arm_avg}
\end{figure*}

In \autoref{fig:arm_avg} we present the value of $B_{\rm pred}$ estimated with the three alternative approaches to estimating $\langle e_{\rm c}\rangle$ (\autoref{eq:Beq_avg}, \autoref{eq:Beq_SFR}, and \autoref{eq:Beq_W}). Similarly to the {\tt R8} model, the prediction of magnetic field at small scales is improved when we use a constant value of $e_{\rm c}$ across the simulation box rather than assuming local equipartition. The power-law scaling between $B_{\rm pred}$ and $B_{\rm avg}$ is similar in all three cases, with $B_{\rm pred} \propto B_{\rm avg}^{0.82-0.84}$, comparable to the relationship when using the simulated value of $e_{\rm c}$ (\autoref{fig:arm_ec}).

Furthermore, while the absolute normalization of $B_{\rm pred}$ changes slightly between the three alternative estimates, all result in relatively good predictions of $B_{\rm avg}$. Across all three alternative estimates, the box-averaged value of $B_{\rm pred}/B_{\rm avg}$ is between 0.6-1.2.

\subsection{Polarization}
\label{sec:pol}

In addition to probing the magnetic field strength, synchrotron observations also provide a way to determine the orientation of magnetic fields when we consider polarized emission. We evaluate the accuracy with which the magnetic field direction can be extracted from the Stokes parameters, $Q_\nu$ and $U_\nu$. To do so, we find the polarization angle, $\varphi$, using \autoref{eq:phi} for each of the {\tt R8} and {\tt R8 Arm} snapshots. We rotate this angle by 90$^\circ$ to recover an estimate of the magnetic field direction. Then, we compare this to the angle made by the true magnetic field in the plane of the sky. We estimate the average angle of the simulated magnetic field as
\begin{equation}
    \varphi_B = \arctan \left(\frac{B_{\rm y, avg}}{B_{\rm x, avg}} \right)
    \label{eq:phi_B}
\end{equation}
where $B_{\rm x, avg}$ and $B_{\rm y, avg}$ are the mass-weighted average x and y components of the magnetic field along the line of sight. Note that we do not include Faraday rotation in our calculation, just polarized emission.\footnote{More sophisticated analyses, making use of the intensity of polarized emission and the rotation measure, can be applied to separate out the strengths of turbulent and regular components of the magnetic field \citep{Beck_25}.} 

\begin{figure}
    \centering
    \includegraphics[width=\linewidth]{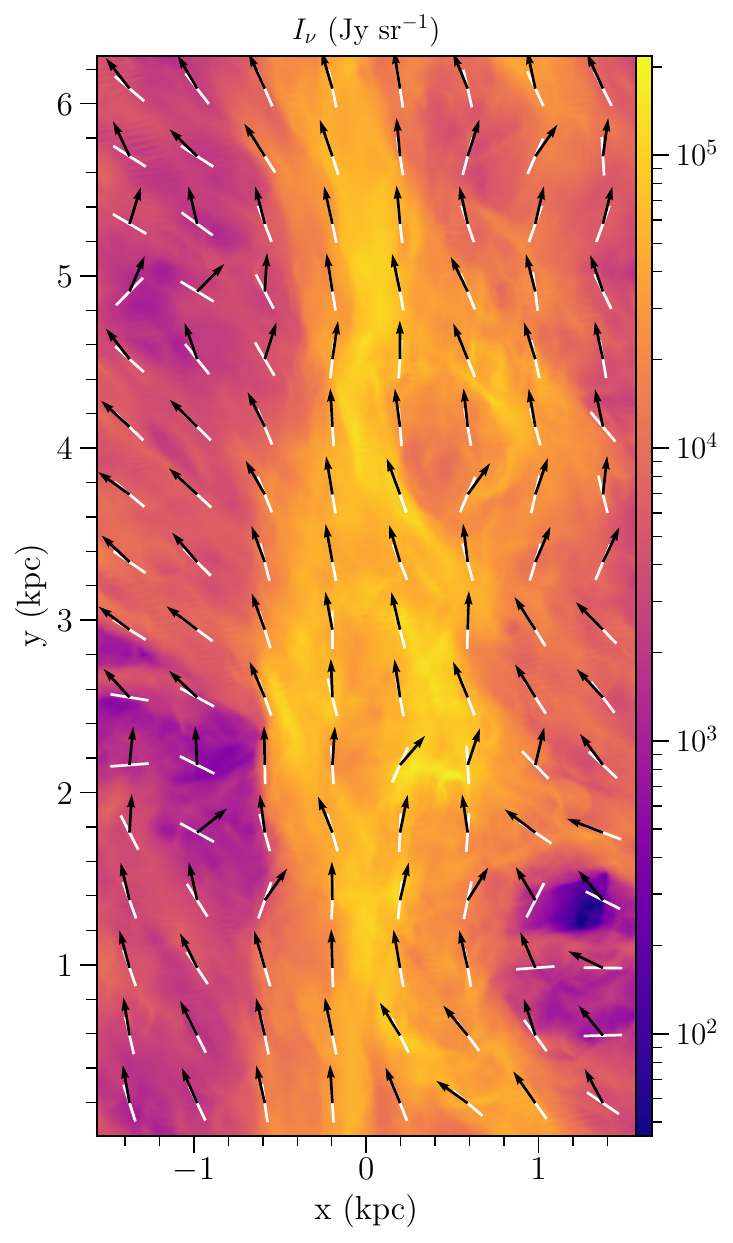}
    \caption{Synchrotron emission at 1.5 GHz from one snapshot of the {\tt R8 Arm} model. The white bars represent the magnetic field direction estimated from polarization measurements. This is given by \autoref{eq:phi} rotated by 90$^\circ$. The black arrows represent the mass-weighted average magnetic field direction projected into the x-y plane (\autoref{eq:phi_B}).}
    \label{fig:pol_snap}
\end{figure}

In \autoref{fig:pol_snap}, we show the synchrotron intensity at 1.5 GHz for one snapshot of the {\tt R8 Arm} model (also shown in \autoref{fig:arm_snap}). Projected onto the snapshot, we show the magnetic field direction (\autoref{eq:phi_B}) with black arrows. We also include the angle of the magnetic field as estimated by the Stokes parameters with white bars. This value is ambiguous up to 180$^\circ$ so does not have a fixed direction. Qualitatively, we see that the magnetic field is generally aligned with the values estimated by polarization, as would be expected. However, the two angles are not strictly aligned, likely due to the variations in the magnetic field and synchrotron emission along the line of sight.

\begin{figure}
    \centering
    \includegraphics[width=\linewidth]{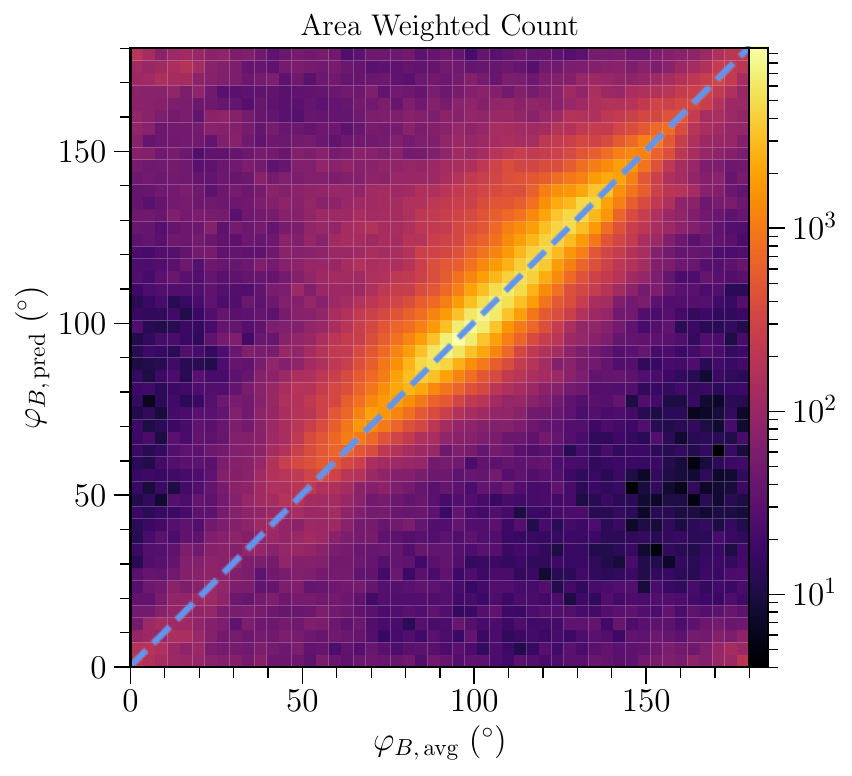}
    \caption{Comparison of the mass-weighted average magnetic field direction projected into the x-y plane ($\varphi_{B, \rm avg}$) to the observationally estimated magnetic field direction $\varphi_{B, \rm pred} = \varphi + 90^\circ$ (\autoref{eq:phi}) including all simulation snapshots. The blue dashed line represents $\varphi_{B, \rm avg} = \varphi_{B, \rm pred}$. Overall, we see good agreement between $\varphi_{B, \rm avg}$ and $\varphi_{B, \rm pred}$.}
    \label{fig:pol_hist}
\end{figure}

In \autoref{fig:pol_hist}, we compare the true direction of the magnetic field ($\varphi_{B, \rm avg}$) to the value estimated using \autoref{eq:phi} ($\varphi_{B, \rm pred}$) across all {\tt R8} and {\tt R8 Arm} snapshots in a two-dimensional histogram. Because the polarization angle is ambiguous up to 180$^\circ$, we limit both angles to between 0 and 180$^\circ$. From the distribution in \autoref{fig:pol_hist}, we see that the observational estimate does a good job of extracting the true magnetic field direction. Quantitatively, in 73\% of the pixels, $\varphi_{B, \rm pred}$ is within 15$^\circ$ of $\varphi_{B, \rm avg}$. If we define the true magnetic field direction using a volume-weighted, rather than mass-weighted, average along the line of sight, we find similar results with 71\% of the pixels having $\varphi_{B, \rm pred}$ within 15$^\circ$ of $\varphi_{B, \rm avg}$.

\section{Discussion}\label{sec:disc}

\subsection{Equipartition at different scales}

In both the {\tt R8} and {\tt R8 Arm} models, we find that the traditional equipartition estimate of magnetic field strength based on synchrotron emission is reasonable (within $\sim 40\%$) when considering the full simulation boxes. This is not surprising, as our simulations show that the magnetic and CR energy densities agree within a factor of two at the midplane (\autoref{sec:measured_e}). However, we do find significant disagreement with the equipartition estimate at the resolution scale of our simulations. This is consistent with expectations, as both observational and theoretical work has predicted that equipartition should not hold below $\sim1$ kpc \citep[e.g.][]{stepanov_observational_2014, seta_revisiting_2019}. In our simulations, this disagreement is due to the fact that the CR energy density is quite uniform across the x-y plane (due to rapid CR diffusion in the neutral gas), while the magnetic field has much more significant variation. 

We note that the for the {\tt R8 Arm} model, if we consider smaller patches of the simulation having the same size as the {\tt R8} model ($\sim1$ kpc), the equipartition estimate fails to recover the true field strength. This is because the CR energy density on the box scale in both simulations is determined by a competition between CR injection by supernovae and loss by advection and diffusion (which is relatively independent of box size), while the uniformity of the CRs in the plane of the disk is a consequence of the lack of any background gradients and rapid CR diffusion. Our simulations cannot therefore directly predict the scale at which approximate equipartition would hold. Gamma-ray observations of our Galaxy show large scale variations in the CR energy density as a function of galactic radius \citep[e.g.][]{Acero2016, yang2016}. It is likely this large-scale spatial gradient that determines the scale at which approximate equipartition is a valid assumption in Milky Way-like galaxies.

\subsection{Comparison to other work}

In general, other investigations that produce synthetic synchrotron emission maps from galaxy simulations and test the magnetic field equipartition estimate find results broadly consistent with what we have presented here. For example, using cosmological zoom-in FIRE simulations \citep{hopkins_first_2022}, \cite{2024MNRAS.52711707P} find that, although the traditional equipartition estimate provides overall a reasonable approximation of the true magnetic field strength, it generally underestimates it. They identify a reason for this mismatch that is consistent with our findings: because CR transport is highly diffusive, the CR distribution is much smoother than that of the magnetic field, which is clumpy and generally stronger in warm/cold neutral gas, leading to violations of equipartition in these regions. \cite{2024MNRAS.52711707P} show that introducing a parameter accounting for magnetic field clumpiness into the equipartition formula yields magnetic field values in better agreement with the simulated ones.

\cite{dacunha_overestimation_2024} produce synthetic synchrotron emission maps of galaxies using  cosmological simulations by \citet{Martin-Alvarez+20, Martin-Alvarez+21}. These simulations, however, do not directly model CR transport. Therefore, \cite{dacunha_overestimation_2024} make simplified assumptions about the spatial and spectral distributions of CREs to produce the maps. They find a much larger discrepancy between the simulated magnetic field strength and the traditional equipartition estimate compared to both our results and those of \cite{2024MNRAS.52711707P}. Similar to our results, however, \cite{dacunha_overestimation_2024} do find that the ratio between the equipartition estimate and the true magnetic field depends on the magnitude of the intrinsic magnetic field. They show that as the simulated magnetic field strength increases, the equipartition estimate is more likely to underestimate rather than overestimate the true magnetic field strength.

The study most directly comparable to ours, in terms of both resolution and CR physics -- specifically, the use of a variable scattering coefficient based on self-confinement -- is the recent work by \citet{chiu2025}, which analyzes a simulation of an isolated Milky Way-like galaxy \citep{Thomas2025}. They conclude that the assumption of energy equipartition is, overall, a reasonable approximation, as it has only a limited impact on the estimated magnetic field value. This is because the ratio between the true and equipartition fields tends to be much closer to unity than that between the magnetic and CR energy densities. This finding is consistent with our results: in our simulations, the true magnetic field and the equipartition estimate differ by less than an order of magnitude, whereas the magnetic and CR energy densities can differ by up to three orders of magnitude (see Figure~22 in \citealt{armillotta_cosmic-ray_2021}). However, when checking the equipartition on a pixel-by-pixel basis, they also find that the equipartition estimate deviates from the true value in regions of strong magnetic field (corresponding to high-density cold/warm gas), where it tends to be underestimated.

\subsection{Accuracy of alternative estimates}
\label{sec:disc_2}

In this work, we presented three alternative approximations for the CR energy density which might be used instead of the traditional equipartition calculation to estimate the magnetic field strength in galaxies. These methods are particularly important for inferring small-scale magnetic field structure from synchrotron observations, where we have shown the traditional equipartition estimate does not accurately predict the magnetic field strength. All three estimates amount to the same modification to the original equipartition estimate, i.e. that the CR energy density is constant on small scales. In each case, $B_{\perp} \propto I_{\nu}^{\frac{1}{\alpha + 1}}$, and the primary difference is in how the proportionality factor (which is a function of large-scale galactic conditions) is determined.

If we knew the scale over which the CR energy density is approximately constant, then we would have an implementable observational model for estimating magnetic field strength from radio observations. We could take the CR energy density to be a constant value using any of the three alternative estimators we proposed in \autoref{sec:alt}. From our simulation results, however, we clearly cannot constrain this length scale, as we find the CR energy density to be constant on the simulation box scale independent of the size of the box. One could plausibly average over a length scale comparable to the local gradient scale of relevant observables in a galactic disk. To determine the appropriate averaging scale more precisely, however, we would need to use results from global models of CR transport.

In all of the extragalactic magnetic field estimators considered here, there is uncertainty in the overall normalization of the predicted magnetic field strength compared to the true value. While all three methods have similar scaling between $B_{\rm pred}$ and $B_{\rm avg}$, they differ in magnitude by up to $\sim40\%$. These estimates, as well as the original equipartition estimate, rely on multiple free parameters to set the magnitude of the predicted magnetic field. For example, all estimates are a function of $K_0$ (the ratio of the CR proton to electron number densities at energies of a few GeV) and $\ell_z$ (the vertical length scale for emission). These parameters may in principle be set either from observational measurements or theory, but are not necessarily fixed across all environments. Two of our alternative estimates in principle capture potential systematic variations in the global ratio of CR energy density to magnetic energy density using fits to simulations. However, these fits are based on a very limited number of simulations, and it is likely that system parameters beyond just the local star formation rate or ISM weight may affect $\langle e_\mathrm{c}\rangle/\langle e_\mathrm{mag}\rangle$. The simulations themselves are also subject to assumptions regarding the fraction of SN energy injected to CRs above 1 GeV. Furthermore, introducing additional observations adds to the uncertainty of the final estimate, although this is not likely to be the dominant source of error. At present, it is premature to conclude if one alternative estimate would perform significantly better when applied to real observations. 

Finally, a caveat is that our tests in this work are based on simulations representing only one environment, i.e. conditions similar to the solar neighborhood. Formally, these results should not be extrapolated to galaxies with significantly different conditions compared to Milky Way-like spirals until we expand our numerical experiments to simulations spanning a wider range of environments. Nonetheless, as we discuss next, there are theoretical reasons suggesting the present results may hold more broadly. 

\subsection{Equipartition via Mutual Inheritance}

For both the real Milky Way galaxy and our simulations, approximate equipartition holds between $e_\mathrm{c}$ and $e_\mathrm{mag}$ on large scales. However, this does not come about in our simulations because of a dynamical interaction between CRs and the background magnetized gas. The original TIGRESS simulations were evolved without CRs, and the CR transport is performed essentially as a post-processing operation. Thus, the approximate equipartition between $e_\mathrm{c}$ and $e_\mathrm{mag}$ is not the result of a causal dynamical relationship between the two ISM components, but because different feedback processes coincidentally yield similar energy density levels. Both $e_\mathrm{c}$ and $e_\mathrm{mag}$ are a response to the energy returned from massive stars as supernovae, so that they both are expected to be proportional to the star formation rate and therefore to each other.  However, very different physics is involved in setting the coefficients.  

As discussed in \autoref{sec:SFR_scaling}, from a theoretical point of view, and as supported by numerical simulations analyzed in \citet{2025arXiv250903519H}, it is expected that $\langle e_\mathrm{c}\rangle = 3 P_\mathrm{c}$ is set from $P_\mathrm{c}=\Upsilon_\mathrm{c} \Sigma_\mathrm{SFR}$ with $\Upsilon_\mathrm{c}\approx\epsilon_\mathrm{c} E_\mathrm{SN}/(v_\mathrm{c,eff} 8 m_\star)$ when losses are neglected.  In the GeV regime, which dominates the integrated CR energy density, the scattering rate is high so that CR transport is controlled by the combination of advection and Alfv\'enic streaming, $v_\mathrm{c,eff} = v + v_{A,i}$, in the extraplanar ionized gas \citep{Armillotta2025}.  The fountain flow velocity of this gas is a response to forces that ultimately can be traced to the thermal pressure gradients in superbubbles created by supernova blasts \citep[see][]{Kim_Ostriker2018,Vijayan2020}.\footnote{Preliminary results from fully self-consistent TIGRESS CR-MHD simulations find similar values of $v_\mathrm{c,eff}$ to those from TIGRESS simulations without CRs, implying that CR pressure gradients do not significantly change gas outflow velocities  (C.-G. Kim et al., in prep).}  The relation $\Upsilon_\mathrm{c} \propto \Sigma_\mathrm{SFR}^{-0.23}$ cited in \autoref{eq:Upsilon_c} is likely due to enhanced spatio-temporal correlation of supernovae at higher $\Sigma_\mathrm{SFR}$ increasing the outflow velocity and therefore $v_\mathrm{c,eff}$ (see also fits for $v_\mathrm{out}$ as a function of $\Sigma_\mathrm{SFR}$ reported in \citealt{kim_first_2020}).  

Magnetic energy in disk galaxies is in approximate equipartition with turbulent energy, which is understandable since magnetic fields are maintained largely by the interaction of turbulence with sheared rotation \citep[e.g.][]{Brandenburg2023}. The turbulence that generates the magnetic field is in turn driven by supernovae, following $P_\mathrm{turb} = \Upsilon_\mathrm{turb} \Sigma_\mathrm{SFR}$ with $\Upsilon_\mathrm{turb}\approx p_*/(4m_*)$ for $p_*$ the momentum injection per supernova \citep[see][and references therein]{ostriker_pressure-regulated_2022}. Physically,  $p_* \approx E_\mathrm{SN}/v_\mathrm{cool}$ for $v_\mathrm{cool}$ the velocity at which supernova shocks become strongly radiative \citep{Kim_Ostriker2015,Kim_2017}. 

Based on the above arguments, if $P_\mathrm{mag}\sim P_\mathrm{turb}$ we would expect $e_\mathrm{c}/e_\mathrm{mag}\sim (3/2) \epsilon_\mathrm{c} v_\mathrm{cool}/v_\mathrm{c,eff}$.  For conditions similar to the solar neighborhood, $v_\mathrm{cool}\sim 250$~km/s, and more generally this increases weakly at higher ambient density and metallicity \citep[see e.g.][]{Kim_JG_2023}, where higher density is also correlated with higher $\Sigma_\mathrm{SFR}$.  For solar neighborhood conditions, $v_\mathrm{c,eff}\sim 15-20$ km/s, and this also appears to increase in regions of higher $\Sigma_\mathrm{SFR}$ \citep{2025arXiv250903519H}. Thus, $e_\mathrm{c}/e_\mathrm{mag}\sim2$ is expected on theoretical grounds for the solar neighborhood.  More generally, we would expect $e_\mathrm{c}/e_\mathrm{mag}=3 \Upsilon_\mathrm{c}/\Upsilon_\mathrm{mag}$, and based on TIGRESS simulations measuring $\Upsilon_\mathrm{mag}$  \citep{Kim_2024} the scaling is expected to vary as $\propto \Sigma_\mathrm{SFR}^{0}$ to $\Sigma_\mathrm{SFR}^{0.2}$.  Thus, the approximate equipartition that holds for the Milky Way may also roughly hold in other galactic environments, subject to potential variations in $\epsilon_\mathrm{c}$.  We emphasize, however, that this should be considered a consequence of mutual inheritance from star formation feedback, rather than magnetic energy setting CR energy (or vice versa).

\section{Summary and Conclusions}\label{sec:conc}

In this work, we use post-processed models of spectrally resolved CRE transport in the TIGRESS MHD simulations of the ISM to test how well the magnetic field strength and direction can be recovered using common observational methods. We find that observations of polarized synchrotron emission are a good probe of the magnetic field direction. The traditional equipartition estimate of magnetic field strength does well at the simulation scale, both for the {\tt R8} and larger {\tt R8 Arm} models, but breaks down if we consider smaller scales. Because we find equipartition to hold at the largest scale independent of the simulation box size, we cannot use the simulations presented in this work to directly constrain an ``equipartition'' scale. To do this, we would require global simulations of CR transport.

We show that the agreement between the predicted and true magnetic field strength at the simulation resolution scale ($\sim 10$pc) can be improved by assuming a constant CR energy density $\langle e_\mathrm{c}\rangle$ across the simulation box. The value of $\langle e_\mathrm{c}\rangle$  can be estimated using the assumption of equipartition at large scales, or by bringing in additional observables such as the SFR or gas weight. These alternative approximations result in equally good estimates of magnetic field strength when considering the distribution of $B_{\rm pred}$ compared to $B_{\rm avg}$. However, each of these alternative observables result in slightly different estimates of $\langle e_\mathrm{c}\rangle$, changing the overall normalization of $B_{\rm pred}$. In the future, results from additional simulations in more environments, including global galactic models, can be brought in to improve calibrations of the relationship between $\langle e_\mathrm{c}\rangle$ and galactic observables.

At present, almost all measurements of magnetic field strength in extragalactic sources rely on the equipartition estimate \citep[see][for just a few examples]{tabatabaei_high-resolution_2008, heesen_cosmic_2009, soida_large_2011, mora_magnetic_2013, mulcahy_modelling_2016, mora-partiarroyo_chang-es_2019, stein_chang-es_2023}. Additionally, future surveys of extragalactic sources with telescopes such as the next-generation VLA \citep[e.g.][]{mckinnon_2019} or Square Kilometre Array \citep[e.g.][]{braun_2015} will reach resolutions comparable to the simulations presented in this work. Based on our results, as well as other theoretical work \citep[see e.g.][]{stepanov_observational_2014, seta_revisiting_2019} we argue that at small scales the traditional equipartition estimate should not be applied directly, but can be easily modified to probe the underlying magnetic field strength.

\begin{acknowledgements}
We are grateful to the referee for a helpful report. Support for this work was provided by grant 510940 from the Simons Foundation to ECO and grant AST-2407119 from the NSF to LA and ECO. LA was supported in part by the INAF Astrophysical fellowship initiative.  EQ was supported in part by NSF AST grant 2107872 and by a Simons Investigator grant. 
    
\end{acknowledgements}

\software{Athena++ \citep{Stone2020}, Athena \citep{stone_athena_2008, stone_gardiner2009}, Matplotlib \citep{Hunter:2007}, NumPy \citep{harris2020array}, SciPy \citep{2020SciPy-NMeth}, Astropy \citep{astropy:2013, astropy:2018, astropy:2022}, xarray \citep{hoyer2017xarray}}

\bibliography{references}
\bibliographystyle{aasjournal}

\end{document}